\documentclass{article}  

\usepackage{amsmath,amssymb,amsfonts}
\newtheorem{theorem}{Theorem}[section]

\newtheorem{definition}[theorem]{Definition}

\title{Conformal Einstein evolution
\footnote{To appear in: J. Frauendiener, H. Friedrich (eds.),
``The Conformal Structure of Spacetime: Geometry, Analysis, Numerics'', 
Springer, Lecture Notes in Physics, Berlin, 2002.}}
\author{Helmut Friedrich\\ 
Max-Planck-Institut f\"ur Gravitationsphysik\\
Am M\"uhlenberg 1\\
14476 Golm\\
Germany}

\begin{document}
\maketitle

\begin{abstract}
We discuss various properties of the conformal field equations and their
consequences for the asymptotic structure of space-times.
\end{abstract}

%\newpage

%%%%%%%%%%%%

\section{Introduction}

The causal structure, which decides whether a signal can be sent between
two space-time points, is of fundamental importance for all of field
theory. In the case of Einstein's theory of gravitation, general
relativity, it acquires a particular significance, because the null cone
structure underlying the causal relations is itself a basic unknown of the
field equations and as such subject to the causal relations in the small.

From the beginning of general relativity the effect of `deflection of
light in gravitational fields', demonstrated nowadays so clearly by
the gravitational lensing phenomena, has been considered theoretically
and observationally as one of the clearest manifestations of the
deviation of the light cone field in general relativity from that of
special relativity. As seen in the article by G. Galloway in this volume,
which shows, among other things, the ubiquity  of caustics on wave fronts
in solutions to Einstein's field equations, it is not only a local
effect but also related to the global structure of space-times. 

This relation to the global structure of space-time becomes quite
pronounced in the case of black holes, which seem to be on the verge now
of also becoming `observed' phenomena. The observational situation as well
as various theoretical aspects of the idea are addressed directly in P.
Chru\'sciel's article in this volume. However, tacitly or explicitly, the
related questions of gravitational collapse, merger of black holes,
generation of gravitational radiation, singularities of gravitational
fields, etc. motivate to a large extent all studies in this volume.

The dual and critical role of the light cone structure in Einstein's
theory and in physics in general led some researchers to speculate that
the null cone structure might be a classical space-time reflection of a
fundamental feature of a quantum theory of gravitation. This motivated
efforts to recast the classical theory into forms which put the null cone
structure, or associated structures like wave fronts, into the
foreground. The article by S. Frittelli, N. Kamran, and E. Newman as well
as the article by R. Penrose in this volume are aiming into this
direction. Trying instead to derive qualitative and quantitative results
about the solutions, the remaining articles explore and exploit, in one
way or other, the light cone structure along more conventional lines.

That in studies of the global structure of gravitational fields the light
cone structure acquires a particular technical interest, has been
demonstrated most impressively by the proofs of the singularity theorems
(\cite{hawking}, \cite{penrose:coll:let}, cf. also \cite{hawking:ellis}).
It provides the most important tool for analyzing space-times in the
large. While these theorems show that under reasonable
assumptions space-times are singular in the sense that they contain
non-complete causal geodesics, they do not provide any information on the
specific nature of the singularities. It is expected that in general
curvature singularities will arise, however, the theorems give not the
slightest hint whether this will be the case, not to mention the causal
nature of the singularities or their strength. 
  
In spite of various attempts to obtain more detailed results, the related
problem of cosmic censorship (cf. \cite{penrose:sing-t-as},
\cite{penrose:chandra}, \cite{wald:cosc}) still poses some of the most
important outstanding questions of classical general relativity. While
the limited use of the field equations made in the singularity theorems
contributes to their generality, the same fact is responsible for the
scarcity of details supplied by them. In view of the enormous richness of
pathological behaviour observed on the level of pure space-time theory
(cf. \cite{ellis:schmidt} and the literature given there) it is not
surprising that attacks at the problem of cosmic censorship which rely
mainly on the topological methods used in the proof of the singularity
theorems met little success. 

To disclose the more subtle secrets of Einstein's theory we need a
deeper understanding of the causal structure in the context of Einstein's
field equations. In any serious investigation of the problem the
consequences of these equations, which represent the principal
restrictive principle of the theory, must be taken into account
in all completeness. Getting control on the evolution process defined by
Einstein's equations still remains the most important technical task in
classical general relativity.

There arises the question of how to combine the analysis of the field
equations with that of the causal structure. In Einstein's theory these
structures are particularly closely related. By the local causality
requirement the causal structure is determined by the null cones of the
metric. Conversely, provided certain pathologies are excluded, the causal
structure can be used to reconstruct the null cone structure. Moreover,
the physical characteristics of the field equations, which largely govern
the evolution process, are given by the null hypersurfaces defined
by the solutions. These tight relations between causal structure, null
cone structure, and characteristics structure clearly offer possibilities
for the desired analysis, but it may be noted that they are also
responsible for the {\it quasi-linearity} of the equations which,
together with their {\it gauge hyperbolicity}, is the main source of our
difficulties. 

The null cone structure is equivalent to the conformal structure, given by
the information which is retained if the metric is subject to arbitrary
conformal rescalings $g \rightarrow \Omega^2\,g$ with conformal factors 
$\Omega > 0$. This may allow us to relate the null cone structure in a
new way to the structure of the field equations and to the large
scale behaviour of their solutions.  

However, to what extent can we draw from the behaviour of the field
equations under conformal rescalings of the metric conclusions on the
large scale behaviour of the solution? The present article is trying to
give some answer to this question. The latter might be 
rather aimless were it not for a concrete proposal pointing precisely
into this direction. Motivated by earlier studies of the important
physical question of how to characterize gravitational radiation without
relying on approximations, R. Penrose suggested to analyses the asymptotic
behaviour of gravitational fields in terms of extensions of the conformal
structure to null infinity (\cite{penrose:scri:let},
\cite{penrose:scri}). 

The subsequent attempt (\cite{friedrich:1981a}, \cite{friedrich:1981b}) to
resolve the controversy concerning the assumptions made in
\cite{penrose:scri:let} and to understand the smooth interplay
between the conformal structure and the field equations suggested in
\cite{penrose:scri} disclosed the very peculiar behaviour of Einstein's
field equations under conformal rescalings. They turned out to be {\it
conformally regular} and assume with this property a very special position
between {\it conformally singular} equations like the massive
Klein-Gordon equations on a given background and {\it conformally
invariant} equations like the Maxwell or Yang-Mills equations in four
dimensions. It will be the purpose of this article to explain the origin,
the properties, and the consequences of the resulting {\it conformal
field equations}. 

Almost all the results described in the following have been obtained by
considering as a guiding problem the question to what extent the
assumptions underlying \cite{penrose:scri:let} can be justified. We are
beginning to see now that this question proved fruitful because these
assumptions refer to the key geometrical structure and draw such a sharp
line between what is true and what isn't that they required, and still
require, a most careful and detailed analysis. The results obtained so far
indicate that in the most interesting case, that of a vanishing
cosmological constant, the situation is much too subtle to simply allow us
to say that the assumptions of \cite{penrose:scri:let} are right or wrong.

In the following we shall make a particular effort to sort out the
delicate issues which arise here, hoping to put a controversy, whose
`arguments' too often relied more on personal taste than on insight,
onto a rational basis. Such an effort is appropriate because due to
the results which are available now the mathematical analysis is
coming close to a point where the more subtle mathematical and
physical questions concerning the foundations of the theory of
isolated self gravitating systems become amenable to detailed
discussions.

The investigations are now way beyond a situation which would allow us
to arrive at reliable statements by investigations of the linearized
equations, `physical arguments' etc. In fact, it is difficult to develop
any intuition on physical grounds for the extreme questions about 
solutions of Einstein's equations which are being asked here. 
Consequently, most results are stated in the form of existence theorems
(referred to in the following only in a non-technical way). The physicist
mainly interested in wave forms and observational facts, who conceives of
this as merely interesting mathematics, may be reminded that abstract
statements about the existence of general classes of solutions with
certain global properties usually only epitomize collections of results
about details of the solutions, particular features of the equations, and
new techniques with many more applications. 

More importantly, it may be pointed out to him that the results alluded to
above become of a practical use now, which was not anticipated in
\cite{penrose:scri:let}. As discussed in the articles by J. Frauendiener
and S. Husa in this volume they form a basis for numerical calculations
covering asymptotic parts of space-time.  Most remarkably, though still
to be explored in numerical work, they open  the unique possibility to
calculate numerically on finite grids maximal, globally hyperbolic,
asymptotically flat solutions to Einstein's field equations, including
their asymptotics in null directions and the radiation field at null
infinity. These solutions will be determined solely and uniquely from
standard Cauchy data (subject to certain, quite weak, conditions near
space-like infinity) and require besides the idealization of asymptotic
flatness no further approximation.

Working out the details of such truly global solutions, either analytical
or with numerical methods, will still require a considerable amount of
work and inventiveness. However, the main point to be made here is that
both possibilities are open now and both approaches have a lot to gain
from each other. Each of them will pose questions which will shed new
light on problems of the other approach and the answers obtained in one of
them will suggest new solutions to the other one. 

A main open problem in this context is concerned with the
question to what extent it will be possible to approach `time-like
infinity' within a numerical calculation if the solution develops
singularities and event horizons. This problem has hardly been touched
yet. Even for  basic solutions like the Schwarzschild or the Kerr solution
a careful and detailed analysis of time-like infinity in the conformal
picture is not available yet.  Some of these questions and also the
related problem of gauge conditions will be addressed at the end of this
article.

It is clear that there is still a long way to go before we will be able
to obtain wave forms of sufficient generality or arrive at insights
which will bring us closer to a resolution of the problem of cosmic
censorship. Nevertheless, it appears that the interaction between
numerical and analytical methods referred to above offers the best
prospects for further progress. 

Any work on general classes of solutions to the Einstein equations is
necessarily of a very technical nature. In the following we shall 
give a survey on certain results and the current analytical work based on
the conformal fields equations which emphasizes the general ideas. With
the exception of the somewhat detailed discussion of the conformal field
equations, technicalities will be kept at a minimum.

\section{The conformal field equations}

The core of Einstein's theory of gravitation is represented by his  
field equations
\begin{equation}
\label{einstequ}
\tilde{R}_{\mu \nu} - \lambda\,\tilde{g}_{\mu \nu} 
= \kappa\,(T_{\mu \nu} - \frac{1}{2}\,T\,\tilde{g}_{\mu \nu}),
\end{equation}
for the metric $\tilde{g}_{\mu \nu}$, where $\lambda$ denotes the
cosmological constant and $T_{\mu\nu}$ the energy momentum tensor. 
Most properties of the gravitational field should follow from these
equations. Those which don't and have to be `put in by hand' 
characterize general features as well as specific properties of the
physical situation to be modeled. For instance, in the case of cosmological
space-times one may assume compact time slices while in the description
of isolated gravitating systems one will require the existence of Cauchy
hypersurfaces with asymptotically flat ends, but in both cases one has the
freedom to prescribe initial data in agreement with these assumptions. 
In the following we will mainly be interested in the second of these
scenarios. 

Equations (\ref{einstequ}) are known to be \index{gauge hyperbolic}
{\it gauge hyperbolic} in the sense that they are hyperbolic if
suitable {\it gauge conditions} are imposed. Expressed in terms of
coordinates $x^{\mu}$ satisfying the harmonicity condition
$\tilde{\nabla}_{\nu} \tilde{\nabla}^{\nu} x^{\mu} = 0$, equations
(\ref{einstequ}) take the form
\begin{equation}
\label{harmredeinstequ}
\tilde{g}^{\alpha \beta} \partial_{\alpha} \partial_{\beta}\,
\tilde{g}_{\mu \nu} =
Q_{\mu \nu}(\tilde{g}, \partial\tilde{g}, T)
\end{equation}
of a system of wave equations for the metric coefficients 
$\tilde{g}_{\mu \nu}$, where the right hand side depends on the derivatives
of these coefficients up to first order and on the energy momentum tensor.

The form (\ref{harmredeinstequ}) reveals the double role played by the
light cone structure or, equivalently, the conformal structure. On the
one hand the conformal structure is part of the unknown since it
determines the metric up to a positive point dependent factor, on the
other hand it defines the characteristic cone 
$\{\xi \in T^*M |\,\tilde{g}^{\alpha \beta}\,\xi_{\alpha}\,\xi_{\beta} =
0\}$ 
of the form (\ref{harmredeinstequ}) of Einstein's equations (for this and
various other general notions and results concerning the field equations we
refer the reader to \cite{friedrich:rendall}). The characteristics
strongly govern the local propagation properties of Einstein's equations. 
In fact, a more detailed argument involving (\ref{harmredeinstequ})
implies that equations (\ref{einstequ}) respect the local causality
requirement of relativity, provided the equations for the matter fields
entering the energy momentum tensor are chosen appropriately. This is
related to the fact that the characteristic cone determines the domains
of uniqueness for equation
(\ref{einstequ}). 

The tight relations between the structure of characteristics, the
causal structure, and the light cone structure or, equivalently, the
conformal structure, all of which depend on the solution at hand,
is a very distinctive feature of Einstein's theory.
However, the theory proposed by Einstein gives no reason to expect that
the field equations would be particularly well behaved under conformal
rescalings 
$\tilde{g}_{\mu \nu} \rightarrow g_{\mu \nu} = \Omega^2\tilde{g}_{\mu
\nu}$. The equations are chosen to determine isometry classes of
solutions and are thus not conformally covariant. Formally, this is seen
from the transformation law of the Ricci tensor, given in $n$
dimensions by
\begin{equation}
\label{RiccimGtraf}
R_{\nu \rho}[g] = \tilde{R}_{\nu \rho}[\tilde{g}] 
-\frac{n - 2}{\Omega}\,\nabla_{\nu}\,\nabla_{\rho}\,\Omega - 
g_{\nu \rho}\,g^{\lambda \delta}\,
\left(\frac{1}{\Omega}\,\nabla_{\lambda}\,\nabla_{\delta}\,\Omega
- \frac{n - 1}{\Omega^2}\,\nabla_{\lambda}\,\Omega\,\nabla_{\delta}\,\Omega
\right),
\end{equation}
where on the right hand side $\nabla$ denotes the connection 
associated with the metric~$g$.  

If Einstein's equation (\ref{einstequ}) is assumed in this relation and
the conformal factor $\Omega$ is considered as a given function on the
solution manifold, we obtain an equation for the conformal metric $g_{\nu
\rho}$. It should be possible to analyse this equation by applying the
techniques known for Einstein's equation, since the principal parts of the
differential operators occurring in both equations, i.e. the parts
containing the derivatives of highest order, are identical. For
our intended application this equation for $g_{\nu \rho}$ has two
deficiencies. The most conspicuous one is the occurrence of factors
$1/\Omega$, which makes the equation singular at the set 
${\cal J}' = \{\Omega = 0\}$. Since this set will indicate some kind of
infinity of the solution space-time, it will be of particular interest to
us. The second problem arises because the function $\Omega$, in fact also
the manifold underlying the solution and the set ${\cal J}'$, are not
given a priori. They are related to the global geometry of the
solution and must be determined jointly with the metric. 

Remarkably, Einstein's equations allow us to resolve these problems. They
turn out to be {\it conformally regular} in the sense that there exist
conformal representations of the equation which do not contain factors
$\Omega^{-1}$ (resp. factors $\Omega$ in the principal part) and for which
the conformal factor is determined by the equations themselves. To discuss
this further we recall a few facts of conformal geometry.

\subsection{Conformal geometry}

A smooth \index{conformal!structure} conformal structure ${\cal C}$ on
a manifold $M$ is given by a family of smooth locally defined metrics
whose domains of definition exhaust $M$ and which are related on the
intersections of these domains by conformal rescalings with positive
conformal factors. A smooth pseudo-Riemannian space $(M, g)$ uniquely
determines a smooth conformal structure on $M$ which we denote by
${\cal C}_{g}$. Using a suitable partition of unity, it can be shown
that any conformal structure can be obtained this way and we shall
often represent a conformal structure in terms of a globally defined
metric in the conformal class.

A Lorentzian conformal structure ${\cal C}$ determines a null cone at each
point of $M$. Conversely, the null cones fix ${\cal C}$ uniquely
(cf. \cite{hawking:ellis}). As a consequence `space- and
time-like' are well defined for ${\cal C}$ and a few further important
space-time structures are implied. A null hypersurface for ${\cal C}$
is a smooth (say) hypersurface which at each of its points is tangent to
the null cone field. The unique directions of tangency can be integrated
to obtain null curves which rule the null hypersurface. Considered as
point sets, these curves coincide with null geodesics of the metrics in
the conformal class and any null geodesic can locally be obtained this
way. Extending the null geodesics generating a null hypersurface in any of
their affine parameterizations we obtain an unique extension. The
resulting set, called a {\it wave front}, will in general not any longer
be a smooth hypersurface, because families of null geodesics tend to
develop caustics, i.e. form envelopes or intersections. The wave fronts
and their caustics are also invariants of the conformal structure.

On the level of local differential geometry the most important invariant of
a conformal structure ${\cal C}_{g}$ is its conformal Weyl tensor
$ C^{\mu}\,_{\nu \lambda \rho}$ (observe the index positions), which is the
trace-free part in the decomposition
\begin{equation}
\label{curvedecomp}
R^{\mu}\,_{\nu \lambda \rho} =
2\,\{g^{\mu}\,_{[\lambda}\,L_{\rho]\nu}  -
\,g_{\nu[\lambda}\,L_{\rho]}\,^{\mu}\} 
+ C^{\mu}\,_{\nu \lambda \rho}, 
\end{equation}
of the curvature tensor of $g$, where the trace parts are represented by
the tensor
\begin{equation}
\label{mLtens}
L_{\mu \nu} = \frac{1}{n - 2}\,\left(R_{\mu \nu}
- \frac{1}{2\,(n - 1)}\,R\,g_{\mu \nu}\right). 
\end{equation}
In terms of the Levi-Civita connection $\nabla$ of $g$ and some
coordinate basis $\partial_{\mu}$ our conventions are such that
$R^{\mu}\,_{\nu  \lambda \rho}\,X^{\nu}
= (\nabla_{\lambda}\,\nabla_{\rho}
- \nabla_{\rho}\,\nabla_{\lambda})\,X^{\mu}$.
We denote by $R_{\mu \nu} = R^{\rho}\,_{\mu  \rho \nu}$ and $R$ the
Ricci tensor and the Ricci scalar of $g$ respectively.

The class of torsion free covariant derivative operators naturally
associated with a conformal class ${\cal C}$ comprises but is larger than
the set of Levi-Civita operators for the metrics representing ${\cal
C}$. A \index{Weyl connection} {\it Weyl connection} for ${\cal C}_{g}$ is a torsion free
connection
$\hat{\nabla}$ on $M$ which respects the conformal class ${\cal C}_{g}$ in
the following sense. For any $C^1$ curve $x(\lambda)$ defined on a
neighbourhood of the origin of $\mathbb{R}$ and any frame field
$e_k$, $k = 1, \ldots, n$, which is parallelly transported along $x(\lambda)$
with respect to $\hat{\nabla}$, there exists a function 
$\Omega_{\lambda} > 0$ along
$x(\lambda)$ such that $g(e_i, e_k)|_{x(\lambda)} = 
\Omega_{\lambda} ^2\,g(e_i, e_k)|_{x(0)}$. 
It follows from this that    
\begin{equation}
\label{wdg}
\hat{\nabla}_{\rho}\,g_{\mu \nu} = - 2\,f_{\rho}\,g_{\mu \nu},
\end{equation}
with some arbitrary 1-form $f$ on $M$ and that the connection coefficients defined by 
$\nabla_{\partial_{\mu}}\,\partial_{\nu} = 
\Gamma_{\mu}\,^{\rho}\,_{\nu}\,\partial_{\rho}$ and
$\hat{\nabla}_{\partial_{\mu}}\,\partial_{\nu} = 
\hat{\Gamma}_{\mu}\,^{\rho}\,_{\nu}\,\partial_{\rho}$ are related by
\begin{equation}
\label{WGtraf}
\hat{\Gamma}_{\mu}\,^{\rho}\,_{\nu} = 
\Gamma_{\mu}\,^{\rho}\,_{\nu} + S(f)_{\mu}\,^{\rho}\,_{\nu},
\quad\mbox{with}\quad
 S(f)_{\mu}\,^{\rho}\,_{\nu} \equiv
\delta^{\rho}\,_{\mu}\,f_{\nu} 
+ \delta^{\rho}\,_{\nu}\,f_{\mu} 
- g_{\mu \nu}\,g^{\rho \lambda}\,f_{\lambda}.
\end{equation}

In the special case where the 1-form $f$ is exact such that it can 
be written on suitable open sets $U$ in the form
\begin{equation}
\label{intf}
f = - \Omega^{-1}\,d\,\Omega,
\end{equation}
with some smooth function $\Omega > 0$, the connection $\hat{\nabla}$
coincides on $U$ with the Levi-Civita connection $\tilde{\nabla}$ of the
metric $\tilde{g}$ satisfying 
$g_{\mu \nu} = \Omega^2\,\tilde{g}_{\mu \nu}$.

The curvature tensors of the connections $\hat{\nabla}$ and
$\nabla$ satisfying (\ref{WGtraf}) are related by 
\begin{multline*}
  \hat{R}^{\mu}\,_{\nu \lambda \rho} - R^{\mu}\,_{\nu \lambda \rho} =
  2\left\{\nabla_{[\lambda}\,S_{\rho]}\,^{\mu}\,_{\nu} +
    S_{\delta}\,^{\mu}\,_{[\lambda}\,S_{\rho]}\,^{\delta}\,_{\nu}\right\}
= 2\left\{g^{\mu}\,_{[\rho}\,\nabla_{\lambda]}\,f_{\nu} \right.\\
+ \left. \nabla_{[\rho}\,f^{\mu}\,g_{\lambda] \nu} -
g^{\mu}\,_{\nu}\,\nabla_{[\rho}\,f_{\lambda]}  - 
g^{\mu}\,_{[\rho}\,f_{\lambda]}\,f_{\nu} +
g_{\nu[\rho}\,f_{\lambda]}\,f^{\mu} +
g^{\mu}\,_{[\rho}\,g_{\lambda]\nu}\,f_{\delta}\,f^{\delta}\right\},
\end{multline*}
were indices are raised or lowered with respect to $g$.
This implies for the Ricci tensor
$\hat{R}_{\nu \rho} = \hat{R}^{\mu}\,_{\nu \mu \rho}$
and the Ricci scalar $\hat{R} = g^{\nu \rho}\,\hat{R}_{\nu \rho}$ 
\[
\hat{R}_{\nu \rho} 
= R_{\nu \rho} 
- (n - 1)\,\nabla_{\rho}\,f_{\nu} 
+ \nabla_{\nu}\,f_{\rho}
+ (n - 2)\,f_{\nu}\,f_{\rho}
- g_{\nu \rho}\,(\nabla_{\lambda}\,f^{\lambda} 
+ (n - 2)\,f_{\lambda}\,f^{\lambda}),
\]
\[
\hat{R} = R 
- 2\,(n - 1)\,\nabla_{\lambda}\,f^{\lambda} 
- (n - 1)\,(n - 2)\,f_{\lambda}\,f^{\lambda}.
\]
Setting here $f = - \Omega^{-1}\,d\,\Omega$, we obtain for the metrics $g$,
$\tilde{g}$ which are related by the rescaling 
$g = \Omega^2\,\tilde{g}$ the relation (\ref{RiccimGtraf}) of their Ricci
tensors, which implies for their Ricci scalars the relation
\begin{equation}
\label{Rscalct}
4\,\frac{n - 1}{n - 2}\,\nabla_{\mu}\,\nabla^{\mu}\,\theta
- R[g]\,\theta = - \tilde{R}[\tilde{g}]\,\theta^{\frac{n + 2}{n - 2}}.
\end{equation}
For later convenience we set here $\theta = \Omega^{- \frac{n - 2}{2}}$.

The analogue of the decomposition (\ref{curvedecomp}) takes the form
\begin{equation}
\label{Wcurvedecomp}
\hat{R}^{\mu}\,_{\nu \lambda \rho} =
2\,\{g^{\mu}\,_{[\lambda}\,\hat{L}_{\rho]\nu}  -
g^{\mu}\,_{\nu}\,\hat{L}_{[\lambda \rho]}  -
\,g_{\nu[\lambda}\,\hat{L}_{\rho]}\,^{\mu}\} 
+ C^{\mu}\,_{\nu \lambda \rho}, 
\end{equation}
where the trace parts of the curvature tensor are represented by the
tensor 
\begin{equation}
\label{WLtens}
\hat{L}_{\mu \nu} = \frac{1}{n - 2}\,\left(\,\hat{R}_{(\mu \nu)}
- \frac{n - 2}{n}\,\hat{R}_{[\mu \nu]}
- \frac{1}{2\,(n - 1)}\,g_{\mu \nu}
\,\hat{R}\,\right). 
\end{equation}
It is related to (\ref{mLtens}) by
\begin{equation}
\label{WLtensrel}
\nabla_{\mu}\,f_{\nu} - f_{\mu}\,f_{\nu}
+ \frac{1}{2}\,g_{\mu \nu}\,f_{\lambda}\,f^{\lambda}
= L_{\mu \nu} - \hat{L}_{\mu \nu}.
\end{equation}
By a straightforward calculation it can be verified 
that (\ref{mLtens}), (\ref{WLtens}) satisfy the
identity
\begin{equation}
\label{Bianchirel}
\hat{\nabla}_{\rho}\,\hat{L}_{\mu \nu} - 
\hat{\nabla}_{\mu}\hat{L}_{\rho \nu} =
\nabla_{\rho}\,L_{\mu \nu} - \nabla_{\mu}\,L_{\rho \nu}
+ f_{\lambda}\,C^{\lambda}\,_{\nu \rho  \mu}.
\end{equation}

\subsubsection{Conformal geodesics}

We have seen that null geodesics, considered as point sets, are
invariants of a Lorentzian conformal structure. With any conformal
structure ${\cal C}_{\tilde{g}}$ is associated in fact a much larger
class of distinguished curves called \index{conformal!geodesics} {\it
conformal geodesics}. We shall only give an outline of their
properties here, referring the reader to \cite{friedrich:i-null},
\cite{friedrich:cg on vac}, and \cite{friedrich:schmidt} for details
and further results. A conformal geodesic curve $x(\tau)$ is obtained,
together with a 1-form $b = b(\tau)$ along the curve, as a solution to
the system of equations
\begin{equation}
\label{acgxequ}
(\tilde{\nabla}_{\dot{x}}\dot{x})^{\mu}
+ S(b)_{\lambda}\,^{\mu}\,_{\rho}\,\dot{x}^{\lambda}\,\dot{x}^{\rho} = 0, 
\end{equation}
\begin{equation}
\label{bcgbequ}
(\tilde{\nabla}_{\dot{x}}b)_{\nu} - \frac{1}{2}\, 
b_{\mu}\,S(b)_{\lambda}\,^{\mu}\,_{\nu}\,\dot{x}^{\lambda} 
= \tilde{L}_{\lambda \nu}\,\dot{x}^{\lambda},
\end{equation}
where $\tilde{L}$ and $S$ are as in (\ref{mLtens}), (\ref{WGtraf}).
There are more conformal geodesics than geodesics for any metric in the
conformal class because for given initial data  
$x_* \in M$, $\dot{x}_* \in T_{x_*} M$, $b_* \in T^*_{x_*} M$
there exists a unique conformal geodesic $x(\tau)$,
$b(\tau)$ near $x_*$ satisfying for given $\tau_* \in \mathbb{R}$
\begin{equation}
\label{cgindat}
x(\tau_*) = x_*,\,\,\,\,\,\,\dot{x}(\tau_*) =
\dot{x}_*,\,\,\,\,\,\,b(\tau_*) = b_*.
\end{equation}
The sign of $\tilde{g}(\dot{x},\dot{x})$ is preserved along a conformal
geodesic but not its modulus. Conformal geodesics, considered as point
sets, are in general different from the metric geodesics of a metric in
the conformal class. Furthermore, they admit general fractional linear
maps as parameter transformations. Initial data with
$\tilde{g}(\dot{x}_*,\dot{x}_*) = 0$,
$\dot{x}_* \neq 0$, determine null curves which coincide, as point sets,
with the null geodesics of the metrics in the conformal class. The fact
that we shall mainly make use of time-like conformal geodesics in the
following, should not be taken as an indications that null or space-like
conformal geodesics can not be useful.

Conformal geodesics are \index{conformal!invariants} {\it conformal
invariants} in the following sense. Denote by $f$ a smooth 1-form
field. Then, if $x(\tau)$, $b(\tau)$ solve the conformal geodesics
equations (\ref{acgxequ}), (\ref{bcgbequ}), the pair $x(\tau)$,
$b(\tau) - f|_{x(\tau)}$ solves the same equations with
$\tilde{\nabla}$ replaced by the connection $\hat{\nabla} =
\tilde{\nabla} + S(f)$ and $L$ by $\hat{L}$, i.e.  $x(\tau)$, in
particular its parameter $\tau$, is independent of the Weyl connection
in the conformal class which is used to write the equations.

Assume that there is given a smooth congruence of conformal geodesics
covering an open set $U$ of our manifold such that the associated 1-forms
$b$ define a smooth field on $U$. Denote by $\hat{\nabla}$ the torsion
free connection on $U$ which defines with the connection entering 
(\ref{acgxequ}), (\ref{bcgbequ}) the difference tensor 
$\hat{\nabla} - \tilde{\nabla} = S(b)$ and denote by $\hat{L}$ the
tensor (\ref{WLtens}) derived from $\hat{\nabla}$. Comparing with 
(\ref{WLtensrel}), we find that equations (\ref{acgxequ}), (\ref{bcgbequ})
can be written 
\begin{equation}
\label{Wacgxequ}
\hat{\nabla}_{\dot{x}}\,\dot{x} = 0, 
\end{equation}
\begin{equation}
\label{Wbcgbequ}
\hat{L}_{\lambda \nu}\,\dot{x}^{\lambda} = 0.
\end{equation}
Let $e_k$ be a frame field satisfying 
\begin{equation}
\label{Wccgxequ}
\hat{\nabla}_{\dot{x}}\,e_k = 0,
\end{equation}
along the congruence, and suppose that $S$ is a hypersurface transverse
to the congruence which meets each of the curves exactly once
and on which $\tilde{g}(e_i, e_k) = \Theta^2_*\eta_{ik}$ with some
function $\Theta_* > 0$.
Following the argument which leads to (\ref{wdg}) we see that  
$\tilde{g}(e_i, e_k) = \Theta^2\,\eta_{ik}$ on $U$ with a function
$\Theta$ satisfying
\begin{equation}
\label{Wdcgxequ}
\hat{\nabla}_{\dot{x}}\,\Theta = \Theta\,<\dot{x}, b>,
\,\,\,\,\,\,
\Theta|_S = \Theta_*.
\end{equation}

\subsection{Derivation of the conformal field equations}

The curvature tensor always satisfies the Bianchi identity
$\nabla_{[\delta}\,R^{\mu}\,_{|\nu|\lambda \rho]} = 0$. Observing
(\ref{curvedecomp}), we can write it in $n \ge 3$ dimensions in the
form
\begin{equation}
\label{fullBian}
\nabla_{[\delta}\,C^{\mu}\,_{|\nu|\lambda \rho]}
= 2\,\{g_{\nu [\lambda}\,\nabla_{\delta}\,L_{\rho]}\,^{\mu}
- g^{\mu}\,_{[\lambda}\,\nabla_{\delta}\,L_{\rho] \nu}\},
\end{equation}
which gives after a contraction
\begin{equation}
\label{1cBian}
\nabla_{\mu}\,C^{\mu}\,_{\nu \lambda \rho}
= (n - 3)\,\{\nabla_{\lambda}\,L_{\rho\nu}
- \nabla_{\rho}\,L_{\lambda\nu}\}.
\end{equation}

We assume now that the {\it physical metric} $\tilde{g}$ satisfies
Einstein's field equation (\ref{einstequ}) and try to express these
equations in terms of the {\it conformal metric}
\begin{equation}
\label{hf:resc}
g = \Omega^2\,\tilde{g},
\end{equation}
and the conformal factor $\Omega > 0$
such that the difficulties pointed out above are avoided. Although
our discussion can be generalized to include certain matter fields
(cf. for instance \cite{friedrich:global}, \cite{huebner:1995} and also,
for an analysis with a different intention and different results, P. Tod,
this volume), we assume for simplicity that matter fields are absent, i.e.
$T_{\mu
\nu} = 0$.  Equation (\ref{1cBian}) reduces then to 
\begin{equation}
\label{v1cBian}
\tilde{\nabla}_{\mu}\,C^{\mu}\,_{\nu \lambda \rho} = 0.
\end{equation}
This equation is of particular interest for us because due to the
identity
\[
\nabla_{\mu}\,(\Omega^{3 - n}\,C^{\mu}\,_{\nu \lambda \rho}) =
\Omega^{3 - n}\,\tilde{\nabla}_{\mu}\,C^{\mu}\,_{\nu \lambda \rho},
\]
which holds for any conformal factor, it has a
certain conformal covariance. It implies for 
\begin{equation}
\label{ddef}
d^{\mu}\,_{\nu \lambda \rho} = \Omega^{3 - n}\,
C^{\mu}\,_{\nu \lambda \rho},
\end{equation}
the \index{Bianchi equation} {\it Bianchi equation} 
\begin{equation}
\label{dequ}
\nabla_{\mu}\,d^{\mu}\,_{\nu \lambda \rho} = 0,
\end{equation}
which, in a sense, represents the core of all the systems to be
considered in the following.
We consider now equation (\ref{curvedecomp}) for the conformal metric $g$
and write it in the form
\begin{equation}
\label{cgequ}
R^{\mu}\,_{\nu \lambda \rho} =
2\,\{g^{\mu}\,_{[\lambda}\,L_{\rho]\nu}  -
\,g_{\nu[\lambda}\,L_{\rho]}\,^{\mu}\} 
+  \Omega^{n-3}\,d^{\mu}\,_{\nu \lambda \rho}.
\end{equation}
Using (\ref{1cBian}) also for $g$ and observing (\ref{ddef}),
(\ref{dequ}), we obtain for $L_{\mu\nu}$, which represents the Ricci
tensor of the conformal metric, the equation 
\begin{equation}
\label{Lequ}
\nabla_{\lambda}\,L_{\rho\nu}
- \nabla_{\rho}\,L_{\lambda\nu} =
\Omega^{n - 4}\,\nabla_{\mu}\Omega\,\,d^{\mu}\,_{\nu \lambda \rho}.
\end{equation}

For later discussion we note how (\ref{cgequ}), (\ref{Lequ}) generalize if
we admit besides rescalings also transitions to Weyl connections which are
related to the Levi-Civita connection
$\nabla$ of $g$ by difference tensors
\begin{equation}
\label{wconn}
\hat{\nabla} - \nabla = S(f),
\end{equation}
with smooth 1-forms $f$. Since  
$\nabla - \tilde{\nabla} = S(\Omega^{-1} \nabla \Omega)$, it follows that
$\hat{\nabla} - \tilde{\nabla} = S(\Omega^{-1}\,d)$ with the smooth
1-form $d_{\mu} = \Omega\,f_{\mu} + \nabla_{\mu} \Omega$. By
(\ref{Wcurvedecomp}) equation (\ref{cgequ}) is then replaced by
\begin{equation}
\label{wconnequ}
\hat{R}^{\mu}\,_{\nu \lambda \rho} =
2\,\{g^{\mu}\,_{[\lambda}\,\hat{L}_{\rho]\nu}  -
g^{\mu}\,_{\nu}\,\hat{L}_{[\lambda \rho]}  -
\,g_{\nu[\lambda}\,\hat{L}_{\rho]}\,^{\mu}\} 
+ \Omega^{n-3}\,d^{\mu}\,_{\nu \lambda \rho},
\end{equation}
while (\ref{Lequ}) and (\ref{Bianchirel}) imply the equation
\begin{equation}
\label{wLequ}
\hat{\nabla}_{\rho}\,\hat{L}_{\mu \nu} - 
\hat{\nabla}_{\mu}\hat{L}_{\rho \nu} 
= \Omega^{n - 4}\,d_{\mu}\,d^{\mu}\,_{\nu \lambda \rho}.
\end{equation}

By admitting rescalings and transitions to Weyl connections we have
artificially introduced a conformal gauge freedom into the equations. If
these equations are to be used to derive information about the physical
metric $\tilde{g}$ by analyzing boundary value problems, this freedom has
to be removed again by imposing gauge conditions such that (i) the
resulting equations ensure that $\tilde{g}$, derived by (\ref{hf:resc}) from
$\Omega$ and $g$, satisfies the field equations, (ii) interesting
information on $\tilde{g}$ can be gained by suitably fixing $\Omega$ and
$f$. Two quite different methods have been discussed so far
(\cite{friedrich:1981a}, \cite{friedrich:1981b}, \cite{friedrich:AdS}).

\subsubsection{The metric conformal field equations}

We consider only conformal rescalings of the metric and try to get
equations which allow us to specify the conformal factor in a useful way.
If we observe (\ref{einstequ}) in the form 
$\tilde{R}_{\mu \nu} = \frac{1}{n}\,\tilde{R}\,\tilde{g}_{\mu\nu}$
in (\ref{RiccimGtraf}), the resulting equation, which we read now as
equation for $\Omega$, can be written in the form 
\begin{equation}
\label{Omegaequ}
\nabla_{\mu}\,\nabla_{\nu}\,\Omega  
= - \Omega\,L_{\mu \nu} 
+ s\,g_{\mu \nu},
\end{equation}
with
\[
s = \frac{1}{n}\,\nabla_{\lambda}\,\nabla^{\lambda}\,\Omega
+ \frac{1}{2\,n\,(n-1)}\,R\,\Omega.
\]
Equation (\ref{Omegaequ}) is obviously overdetermined. If we apply
$\nabla_{\rho}$ to both sides, commute $\nabla_{\rho}$ with $\nabla_{\mu}$,
contract over $\rho$ and $\nu$, and use the twice contracted Bianchi
identity, we obtain the integrability condition
\begin{equation}
\label{sequ}
\nabla_{\mu}\,s = - \nabla^{\nu}\,\Omega\,L_{\nu\mu},
\end{equation}
which we read as an equation for $s$. 

To see the role of the cosmological constant, we observe that the
transformation law of the Ricci scalar gives
\begin{equation}
\label{lambdaequ}
\lambda = (n - 1)\,(2\,\Omega\,s
- \nabla_{\rho}\,\Omega\,\nabla^{\rho}\,\Omega).
\end{equation}
Equations (\ref{Omegaequ}), (\ref{sequ}) then imply for the quantity on
the right hand side 
\[
\frac{1}{2}\,\nabla_{\lambda}\,(2\,\Omega\,s
- \nabla_{\rho}\,\Omega\,\nabla^{\rho}\,\Omega)
= 
\nabla_{\lambda}\,\Omega\,s
+ \Omega\,\nabla_{\lambda}\,s
- \nabla^{\rho}\,\Omega\,\nabla_{\rho}\,\nabla_{\lambda}\,\Omega
\]
\[
=  \nabla_{\lambda}\,\Omega\,s
+ \Omega\,(- \nabla^{\nu}\,\Omega\,L_{\nu\lambda})
- \nabla^{\rho}\,\Omega\,(- \Omega\,L_{\rho \lambda} 
+ g_{\rho \lambda}\,s)
= 0,
\]
which reflects the well known fact that the twice contracted Bianchi
identity implies that $\lambda$ must be constant. It follows that
(\ref{lambdaequ}) will be a consequence of the other equations if it is
arranged to hold at one point by a suitable choice of the initial data.

We consider (\ref{dequ}), (\ref{cgequ}), (\ref{Lequ}), (\ref{Omegaequ}),
(\ref{sequ}), (\ref{lambdaequ}) as equations for the unknown tensor fields
\[
g_{\mu\nu},\,\,\Omega,\,\,s,\,L_{\mu\nu},\,\,
d^{\mu}\,_{\nu \lambda \rho}, 
\]
and refer to them as the \index{conformal!field~equations!metric} {\it metric conformal field equations}. 
These equations hold for any vacuum solution $\tilde{g}$ and any
conformal factor $\Omega$.

A way to impose a gauge condition on the conformal factor is suggested by
the following observation. If $g$ is any metric and $R^*$ a given
function, the metric 
$g^* = \theta^{\frac{4}{n - 2}}\,g$ will have by (\ref{Rscalct}) the Ricci
scalar
$R[g^*] = R^*$, if $\theta$ solves the equation
\[
4\,\frac{n - 1}{n - 2}\,\nabla_{\mu}\,\nabla^{\mu}\,\theta
- R[g]\,\theta = - R^*\,\theta^{\frac{n + 2}{n - 2}}.
\]
Since locally this equation can always be solved for a positive
function $\theta$, it follows that locally the Ricci scalar $R$ of the
conformal metric can be prescribed in the metric conformal field
equations as a \index{gauge!source function} {\it gauge source
function} for the conformal scaling. Its choice will depend on the
given situation and the choices of the other gauge dependent
quantities. We can set $R = 0$ but, depending on the circumstances,
there may be better choices. In any case this leaves the freedom to
specify suitable boundary data for the conformal factor.
 
From the way we obtained the metric conformal field equations it is clear
that they are equivalent to the vacuum field equations (ignoring questions
of smoothness) where $\Omega$ is positive but that they make sense also
where $\Omega$ vanishes. While the system is regular in this sense, it is
also complicated and highly overdetermined. We shall see below that the
formal regularity of the equations does not necessarily imply that the
data must be regular nor does it ensure by itself that the asymptotic
behaviour of the solutions can be controlled conveniently. To get an
insight into this one needs to work out suitable initial value problems
in great detail and control the {\it constraints} as well as the {\it
evolution}. We now take a look at the various parts of the system.

\subsubsection{The Bianchi equation}

For a solution $\tilde{g}$ to the vacuum field equations the Bianchi
identity (\ref{fullBian}) takes the form
\begin{equation}
\label{vfullBian} 
\tilde{\nabla}_{\delta}\,C^{\mu}\,_{\nu \lambda \rho} 
+ \tilde{\nabla}_{\rho}\,C^{\mu}\,_{\nu \delta \lambda} 
+ \tilde{\nabla}_{\lambda}\,C^{\mu}\,_{\nu \rho \delta} 
= 0.
\end{equation}
Contracting with $\tilde{\nabla}_{\delta}$, commuting derivatives, and
observing (\ref{v1cBian}), we obtain
\[
\tilde{\nabla}_{\delta}\,
\tilde{\nabla}^{\delta}\,C^{\mu}\,_{\nu \tau \rho} =
2\,\{
C^{\mu}\,_{\pi \delta \rho}\,C^{\pi}\,_{\nu \tau}\,^{\delta}
+ C^{\mu}\,_{\pi \tau \delta}\,C^{\pi}\,_{\nu \rho}\,^{\delta}
+ C^{\mu}\,_{ \nu \pi \delta}\,C^{\pi}\,_{\tau  \rho}\,^{\delta}
\}
+ 2\,\lambda\,C^{\mu}\,_{\nu \tau \rho}.
\]
Thus, in the Lorentzian case, which will be considered in the following,
we find the well known fact that the Bianchi identity implies a system of
hyperbolic equations for the conformal Weyl tensor in any dimension $n \ge
4$.

In $n = 4$ dimensions we can define the left and right dual of the
conformal Weyl tensor, given by
$^*C_{\mu\nu \tau \rho} = \frac{1}{2}\,
\tilde{\epsilon}_{\mu\nu}\,^{\delta \pi}\,
C_{\delta \pi \tau \rho}$ and 
$C^*_{\mu\nu \tau \rho} = \frac{1}{2}\,
\tilde{\epsilon}_{\tau \rho}\,^{\delta\pi}
\,C_{\mu\nu \delta \pi}$ respectively, and it is well known that they
coincide. It follows that
\[
\tilde{\epsilon}_{\pi}\,^{\delta \tau \rho}
\tilde{\nabla}_{\delta}\,C^{\mu}\,_{\nu \tau \rho}
= 2\,\tilde{\nabla}_{\delta}\,C^{*\mu}\,_{\nu \pi}\,^{\delta}
= 2\,\tilde{\nabla}_{\delta}\,^*C^{\mu}\,_{\nu \pi}\,^{\delta}
=
- \,\tilde{\epsilon}^{\mu}\,_{\nu}\,^{\tau\rho}\,
\tilde{\nabla}_{\delta}\,C^{\delta}\,_{\pi \tau \rho},
\]
which implies that the contracted Bianchi identity (\ref{v1cBian}) is
equivalent to the Bianchi identity (\ref{vfullBian}). It follows that 
(\ref{v1cBian}) and, by the same argument equation (\ref{dequ}), imply
hyperbolic equations.

In dimensions $n \ge 5$ the situation is quite different as can be
seen by a simple function counting. The conformal Weyl tensor has in $n$
dimensions
\[
\frac{1}{12}\,n^2\,(n^2-1) - \frac{1}{2}\,n\,(n+1),
\] 
independent components while the contracted Bianchi identity
(\ref{v1cBian}) provides
\[
\frac{1}{3}\,n\,(n - 2)\,(n + 2),
\]
independent equations. Moreover, if we express (\ref{v1cBian}) in Gauss
coordinates with $x^0$ the affine parameter on the geodesics
orthogonal to the space-like hypersurfaces $\{x^0 = \mathrm{const}.\}$, it follows
that the equations
$\tilde{\nabla}_{\delta}\,C^{\delta}\,_{0 \lambda \rho} = 0$ 
involve only derivatives in the directions
$\partial_{x^1}$, $\ldots$, $\partial_{x^{n - 1}}$ and thus provide
\[
\frac{1}{2}\,n\,(n - 1)
\]
constraints. Consequently, there are not enough equations 
in (\ref{v1cBian}) if $n \ge 6$
while for $n = 5$ there are as many equations as unknowns but not
sufficiently many evolution equations. 
{\it It follows that a system of conformal equations which includes as
equation for the rescaled conformal Weyl tensor only the Bianchi
equation provides a good evolution system only in four dimensions}. 

If we use in $n \ge 5$ dimensions the full Bianchi identities to derive a
system of wave equations for $d^{\mu}\,_{\nu \lambda \rho}$, it appears
that not all terms of the form $\Omega^{-1}\,\nabla_{\mu}\,\Omega$ can be
removed from the equations. It would be interesting to know whether there
exist other conformally covariant equations which could replace the
Bianchi equation in higher dimensions. If it turns out that a conformally
regular representation for the Einstein equation can only be obtained in
four dimensions, it would also be interesting to know whether there is a
principal difference in the asymptotic behaviour of the fields in four
and higher dimensional space-times.

{\it We shall restrict our discussion in the following to the case 
of four dimensions}.

Instead of expressing the intrinsic hyperbolicity of the Bianchi equation
in terms of a system of wave equations we work directly with the first
order system. The benefits of this will be seen later.  There are known
two quite different possibilities to make the hyperbolicity of the
equations manifest, both of them have been used in numerical calculations
(cf. J. Frauendiener and S. Husa, this volume). It has been shown in
\cite{friedrich:2hyp red} how to obtain hyperbolic equations from the
Bianchi equation in the ADM representation of the metric. The other
method, which is based on a frame formalism, has many advantages and can
be adapted to various geometrical situations. We can use real orthonormal
frames (cf. \cite{friedrich:nagy}) but most convenient are
pseudo-orthonormal frames or spin frames. In the spin frame formalism
the equations split almost automatically into constraints and (symmetric
hyperbolic) evolution equations  (cf. \cite{friedrich:1981a},
\cite{friedrich:1hyp red}, \cite{friedrich:global}). We shall use this
fact here to discuss certain features of the Bianchi equation. 

In the spin frame formalism we have a frame field $e_{aa'}$, 
$a, a' = 0, 1$, with $e_{00'}$, $e_{11'}$ real null vector fields and
$e_{01'}$, $e_{10'}$ complex vector fields which are conjugate to each
other. Their only non-vanishing scalar products are given by
$g(e_{00'},e_{11'}) = 1$, $g(e_{01'},e_{10'}) = -1$. In the following
discussion it is useful to think  of these vector fields as being
obtained in the form  
\[
e_{00'} = \frac{1}{\sqrt{2}}\,(e_0 + e_3),\,\,\,
e_{11'} = \frac{1}{\sqrt{2}}\,(e_0 - e_3),
\]
\[
e_{01'} = \frac{1}{\sqrt{2}}\,(e_1 + i\,e_2),\,\,\,
e_{10'} = \frac{1}{\sqrt{2}}\,(e_1 - i\,e_2),
\]
from a frame field $e_k$, $k = 0, \ldots, 3$, satisfying 
$g(e_i, e_k) = \eta_{ik}$. 

The rescaled conformal Weyl tensor is represented by a completely
symmetric spinor field $\phi_{abcd}$, and the Bianchi equation by the
equation
\begin{equation}
\label{scvBian}
0 = \Lambda_{abca'} \equiv \nabla^f\,_{a'}\,\phi_{abcf},
\end{equation}
where $\nabla_{aa'}$ denotes the covariant derivative on spinor fields in
the direction of $e_{aa'}$ and the index is raised with the antisymmetric
spinor field $\epsilon^{ab}$ which satisfies $\epsilon^{01} = 1$. 

If we set $\Lambda_{abcd} = \Lambda_{abca'}\,\tau_d\,^{a'}$ with 
$\tau_a\,^{a'} = \delta_a\,^1\,\delta_{0'}\,^{a'}
- \delta_a\,^0\,\delta_{1'}\,^{a'}$ we get the decomposition
$\Lambda_{abcd} = \Lambda_{(abcd)} 
- \frac{3}{4}\,\epsilon_{d(c}\,\Lambda_{ab)f}\,^f$ into its irreducible
parts
\[
- 2\,\Lambda_{(abcd)} = P\,\phi_{abcd} 
- 2\,{\cal D}_{(d}\,^f\,\phi_{abc)f},\,\,\,\,\,\,\,\,\,\,\,\,
\Lambda_{abf}\,^f = {\cal D}^{ef}\,\phi_{abef}.
\]
Here $P = \tau^{aa'}\,\nabla_{aa'} = \sqrt{2}\,\nabla_{e_0}$ and 
${\cal D}_{ab} = \tau_{(a}\,^{a'}\,\nabla_{b)a'}$ denote covariant 
directional derivatives such that
${\cal D}_{00} = - \nabla_{01'}$, 
${\cal D}_{01} = {\cal D}_{10} = \frac{1}{\sqrt{2}}\,\nabla_{e_3}$,
${\cal D}_{11} = \nabla_{10'}$ 
(cf. \cite{friedrich:global}, \cite{friedrich:AdS} for more details of
the underlying space-spinor formalism). 

In a Cauchy problem one will in general assume $e_0$ to be the future
directed normal to the initial hypersurface $S$. The operators 
${\cal D}_{ab}$ then involve only differentiation in directions tangent to
$S$ and the equations $\Lambda_{abf}\,^f = 0$ are interior equations on
$S$. They represent the six real constraint equations implied on $S$ by
the Bianchi equation. 

The remaining equations for $\phi_{abcd}$ should be evolution equations.
Multiplying the equations by suitable binomial coefficients (and
considering the frame field and the connection coefficients as given), we
find that the system
\begin{equation}
\label{sysyhy}
- \binom{4}{a+b+c+d}\,\Lambda_{(abcd)} = 0,
\end{equation}
is in fact \index{hyperbolic!symmetric} {\it symmetric hyperbolic}, i.e. it has the form
\[
A^{\mu}\,\partial_{\mu}\,u = H(x,u),
\]
with a $\mathbb{C}^N$-valued unknown $u$, a $\mathbb{C}^N$-valued
function $H(x, u)$, and 
$N \times N$-matrix-valued, possibly
$u$-dependent functions
$A^{\mu}$ which are hermitian, i.e. $^T\bar{A}^{\mu} = A^{\mu}$, and for
which there exists at each point $x$ a covector $\xi_{\mu}$ such that
$A^{\mu}\xi_{\mu}$ is positive definite. There exists a well developed
theory for such systems (cf. \cite{friedrichs}, \cite{kato},
\cite{taylor}, and also H.-O. Kreiss and O. Ortiz, this volume). 

While the constraints implied on a given space-like hypersurface are
determined uniquely, there is a large freedom to select useful evolution
systems. In the present case it turns out that any system of the form
\[
0 = 2\,a\,\Lambda_{(0001')},
\]
\[
0 = (c - d)\,\Lambda_{(0011')} - 2\,a\,\Lambda_{(0000')},
\]
\[
0 = (c + d)\,\Lambda_{(0111')} - (c - d)\,\Lambda_{(0010')},
\]
\[
0 = 2\,e\,\Lambda_{(1111')} - (c + d)\,\Lambda_{(0110')},
\]
\[
0 = - 2\,e\,\Lambda_{(1110')},
\]
with $a, c, e > 0$ and $ - (2\,e + c) < d < 2\,a + c$, is symmetric
hyperbolic (the system (\ref{sysyhy}) occurs here as a special
case). This flexibility allows the Bianchi equation to be adapted to
various geometrical situations such as characteristic initial value
problems (\cite{friedrich:1981a}, \cite{friedrich:1981b},
\cite{friedrich:1982}, \cite{friedrich:pure rad}, \cite{kannar:char}),
standard Cauchy problems (\cite{friedrich:n-geod},
\cite{friedrich:1deSitter}, \cite{friedrich:global}), and initial
boundary value problems (\cite{friedrich:AdS},
\cite{friedrich:nagy}). Note that for these systems the set of
\index{characteristic!hypersurfaces} {\it characteristics}, given by the
hypersurfaces of the form $\{f = 0, df \neq 0\}$ with $f \in C^{1}(M)$
on which $\det(A^{\mu}\,f_{,\mu}) = 0$ is satisfied at each of their
points, does not only comprise the null hypersurfaces and can in fact
be quite complicated.

\subsubsection{The constraints}

\index{constraint!equations|(}
Related to the gauge freedom is the fact that, written in suitably adapted
coordinates (and frame field, if used), Einstein's equations as well as
the metric conformal field equations imply inner equations, i.e.
constraints, on any time-like, null, or space-like hypersurface. We will
consider in the following the latter two cases, because they are
particularly important in initial value problems. (Time-like hypersurfaces
are of interest in initial-boundary value problems, however, even in
general analytical discussions of this problem a detailed discussion of
the constraints on these hypersurfaces is not necessarily required
\cite{friedrich:nagy}.)

Null hypersurfaces representing the \index{characteristics!physical}
{\it physical characteristics} of the Einstein equations, the system of
inner equations is quite different in nature from those arising in the
non-null case. There are on the one hand the intrinsic equations
arising because of the gauge freedom and on the other hand the
characteristic inner equations arising due to the intrinsic wave
equation character of the field equations and these equations combine
in a very peculiar way. Let us consider a situation were data are to
be prescribed on a null hypersurface $N$ and another hypersurface
which has a space-like intersection $\Sigma$ with $N$. Then some of
the unknowns can be prescribed {\it completely freely} on $N$. All the
remaining unknowns can then be calculated on $N$ in suitably adapted
coordinates from certain data on $\Sigma$ by solving a certain
hierarchy of inner equations on $N$, each member of which reduces to a
system of ODEs along the null generators at their subsequent
integration. Those equations which are not used in this procedure will
then be satisfied on $N$ if they hold on $\Sigma$ (cf.
\cite{sachs:charprob} for the vacuum case).

The enormous simplification of the equations resulting from these facts
in a suitably adapted gauge and the possibility to integrate the equations
along characteristics `out to infinity' led a number of groups to develop
very efficient {\it characteristic numerical computer codes}  
(cf. R. Bartnik and A. Norton, J. Font, L. Lehner, this volume). 

Though being more complicated, the inner equations induced on null
hypersurfaces by the metric conformal field equations have a similar
structure and, besides the data required to fix the conformal gauge, the
free data are, of course, up to a rescaling the same
(\cite{friedrich:1981b}). 

In the following we will mainly be concerned with Cauchy problems
and therefore consider now space-like initial hypersurfaces.
The standard constraints induced on space-like hypersurfaces by the
vacuum field equations form a system of underdetermined elliptic
quasi-linear equations for the data induced by the 4-metric on
the initial hypersurface, i.e. the inner metric $\tilde{h}_{\alpha
\beta}$ and the extrinsic curvature $\tilde{\chi}_{\alpha \beta}$. There
exists a detailed theory for these equations which, under the assumption
of constant mean extrinsic curvature, reduces the problem of solving the
constraints to a problem of solving a linear elliptic system and a
semi-linear elliptic scalar equation (cf.
\cite{chouqet-bruhat:york}, \cite{chouqet-bruhat:isenberg:york}). 

The metric conformal field equations imply a system of
\index{conformal!constraints} {\it conformal constraints}
(\cite{friedrich:hypivp}, \cite{friedrich:AdS}) on a space-like
hypersurface $S$ which comprises the usual vacuum constraints but
which is more complicated. This is not only due to the artificial
scaling freedom which has been introduced into the vacuum equations
but also because the equations include integrability conditions and
the Gauss-Codazzi equations. There appear to be two ways of dealing
with these equations. The first method solves essentially the standard
vacuum constraints and calculates from the vacuum data the conformal
data after introducing a suitable conformal rescaling. This method,
which has been analyzed with a certain completeness and underlies the
present numerical calculations, introduces certain numerical
difficulties (cf. J.  Frauendiener and S. Husa, this volume). A second
possibility would be to solve the conformal constraints
directly. First steps in a systematic investigation of this approach
are reported in the article by A. Butscher in this volume (cf. also
the discussion in \cite{friedrich:reg constr}).

Any of these approaches has to deal with a problem which is not considered
in the usual discussions of the standard vacuum constraints. In the
situations which are most interesting for our applications the initial
hypersurface contains a subset $\Sigma$, not belonging to the physical
space-time, where the conformal factor vanishes. As seen in detail in the
next sections, this set can be a 2-dimensional boundary
which represents points at null infinity (cf. L. Andersson, this volume)
or  consist of a number of discrete compactification points, 
$\Sigma = \{i_1, i_2, \ldots i_N\}$, representing
space-like infinities (cf. S. Dain, this volume). In principle, $\Sigma$
could consist of a union of such sets or of more complicated sets. In
any of these situations there arises the question about the precise
smoothness of the data near $\Sigma$ in suitably chosen coordinates. 

It turns out that in general, even under the
strongest smoothness assumptions on the `free' data, there may occur
terms of the form $r^k\,\log^lr$ near $\Sigma$, where $r$ denotes the
(non-physical) distance from $\Sigma$.   However, it also turns out that
under mild additional assumptions  logarithmic terms will be absent near
$\Sigma$. We will discuss some of the subtle questions arising here
later. 
\index{constraint!equations|)}

\subsubsection{Gauge conditions, hyperbolicity, and local evolution}

In principle everything known about Einstein's equations also applies
to our system. If we introduce a fictitious energy-momentum tensor by
writing 
$\kappa\,T_{\mu\nu} = 2\,(L_{\mu\nu} - L\,g_{\mu\nu})$ with a constant
$\kappa \neq 0$ (introduced here to emphasize the similarity, we could set
$\kappa = 1$), our equations imply after a contraction of 
(\ref{cgequ}) the system
\[
R_{\mu\nu} - \frac{1}{2}\,R\,g_{\mu\nu} = \kappa\,T_{\mu\nu},
\]  
\[
\nabla_{[\lambda} T_{\rho] \nu} 
- \frac{1}{3}\,\nabla_{[\lambda}T\,g_{\rho]\nu}
= \frac{1}{\kappa}\,\nabla_{\mu}\Omega\,d^{\mu}\,_{\nu\lambda\rho},
\]
\[
\nabla_{\mu}\,d^{\mu}\,_{\nu\lambda\rho} = 0,
\]
\[
\nabla_{\mu}\,\nabla_{\nu} \Omega = - 
\frac{\kappa}{2}\,(T_{\mu\nu} - \frac{1}{3}\,T\,g_{\mu\nu})\,\Omega 
+ s\,g_{\mu\nu}, 
\]
\[
\nabla_{\mu}\,s = -
\frac{\kappa}{2}\,(T_{\mu\nu} - \frac{1}{3}\,T\,g_{\mu\nu}) 
\,\nabla^{\nu}\,\Omega, 
\]
which can be read as Einstein equations with certain matter fields, in
which $R = - \kappa\,T$ represents a gauge source function. We note that
in boundary value problems for these equations it has to be made sure, by
the way the problems are posed and the data are given, that equation
(\ref{cgequ}), which gives geometrical meaning to the fields $L_{\mu\nu}$
and $d^{\mu}\,_{\nu\lambda\rho}$, will be satisfied.

Following the discussion of the Bianchi equation above (or that of
\cite{friedrich:2hyp red}, if one prefers to consider the metric instead
of a frame field as a basic variable), it is not difficult to extract
symmetric hyperbolic evolution equations, the 
\index{reduced equations} {\it reduced equations},
from the last four subsystems. Of course, any choice here will leave us
with constraint equations. It remains to discuss the first equation. In
principle one can bring to bear here all the methods developed for the
ADM equations if one feels most comfortable with these equations.
However, it should be kept in mind that this is only one of the many
possibilities of dealing with equation (\ref{cgequ}). In the references
given in this article there can be found many others which have been
employed successfully under various geometrical assumptions.

If one is dealing with the Einstein equation above or directly with
equation (\ref{cgequ}), in either case one will have to make the
decision whether to use the metric coefficients $g_{\mu\nu}$ or frame
coefficients $e^{\mu}\,_k$ as basic variables. Furthermore, one needs
to choose gauge conditions, i.e. a way to specify the coordinates and,
in the second case, the frame field. Usually this is done in terms of
a \index{gauge!system} {\it gauge system} of differential equations
and suitable initial conditions. A large set of possibilities is known
by now.

Basic objectives in choosing the gauge are the following  (cf. also
\cite{friedrich:rendall}): (i) The gauge should be adapted to the type
of problem (Cauchy problem, initial-boundary value problem,
characteristic initial value problem),  (ii) The evolution equations
should be simple, manifestly hyperbolic, or of some other form (e.g.
`elliptic-hyperbolic') which allows us to handle them in existence
proofs or in numerical calculations (cf. the discussion by H.-O. Kreiss
and O. Ortiz, this volume).  There are available gauge condition which are
governed by implicit wave equations like the harmonic gauge or the
conditions for frame or spin frame fields employed in
\cite{friedrich:1hyp red}, \cite{friedrich:global},
and there are known explicit conditions, in particular
certain geometric gauge conditions, which lead to a reduction of the
number of the unknowns (cf. \cite{friedrich:AdS}, 
\cite{friedrich:2hyp red}  and the discussion below) and which allow us
to control to some extent the development of a degeneracy of the gauge.
(iii) The gauge conditions as well as the constraints should be preserved
during the evolution. (iv) The gauge should allow us to control its
behaviour during the evolution in numerical or in analytical studies.
(v) It should have a sufficiently long `life time'.

While (ii) and (iii) are in the case of hyperbolic equations local
properties which are accessible to direct calculations, (iv) and (v)
depend very much on the given situation and cannot be handled in a
general way. These points belong to the most difficult ones in the large
scale evolution problem in numerical as well as in analytical studies. 

There are known now various ways to perform
\index{hyperbolic!reductions} {\it hyperbolic reductions}
(cf. \cite{friedrich:2hyp red}) for initial value problems for the
metric conformal field equations: given a suitable solution to the
constraints, we find a {\it reduced system} of hyperbolic evolution
equations for which the local existence and uniqueness of solutions
follow from standard theorems, and, by invoking the available
integrability conditions, we can derive a \index{system!subsidiary}
{\it subsidiary system} which allows us to show that all constraints
and gauge conditions are preserved during the evolution if they are
satisfied initially. {\it The significance of the conformal field
equations is determined to a large extent by the fact that they admit
hyperbolic reductions irrespective of the sign of the conformal
factor}.

After we use the large freedom to choose the gauge system and the
reduced system, the subsidiary system is essentially fixed. 
For the analytical discussions of solutions, the latter is only required
to possess the uniqueness property, which ensures that trivial data imply
the trivial solution. However, since in the numerical treatment of the
Einstein equations the constraints are only satisfied up to a numerical
error, the detailed propagation properties of the subsidiary system may
become important in that context. A systematic analysis of how changes
in the gauge system and the reduced system affect the propagation
properties of the subsidiary system and whether this dependence can be
used to choose reduced systems with improved evolution of the
constraints has not been given yet.

\subsubsection{The general conformal field equations}

We will now consider a system of conformal field equations which employs
the additional freedom gained by admitting general Weyl connections.
Combined with suitable gauge conditions it will lead to a system of
evolution equations which is quite different from and in fact much
simpler than the evolution systems obtained from the metric conformal
field equations. 

Assume that $\tilde{g}$ solves Einstein's vacuum field equations with
cosmological constant $\lambda$ and let $g = \Theta^2\,\tilde{g}$ be
given a conformal metric where $\Theta$ is some conformal factor. 
Furthermore, let $b$ be a smooth 1-form. We denote by
$\tilde{\nabla}$ and $\nabla$ the Levi-Civita connection of $\tilde{g}$
and $g$ respectively and by $\hat{\nabla}$ the Weyl connection for
$\tilde{g}$ which satisfies  $\hat{\nabla} - \tilde{\nabla} = S(b)$. 
It follows that $\hat{\nabla} = \nabla + S(f)$ with
$f = b - \Theta^{-1}\, \nabla \Theta$. We define a further 1-form by
setting $d_{\mu} = \Theta\,f_{\mu} + \nabla_{\mu} \Theta$.

Let $e_k$ be a frame field satisfying
$g(e_i, e_k) = \eta_{ik}$. We denote by $\nabla_k$ and $\hat{\nabla}_k$
the covariant derivative in the direction of $e_k$ 
with respect to $\nabla$ and $\hat{\nabla}$
and define the connection coefficients $\hat{\Gamma}_i\,^j\,_k$
of $\hat{\nabla}$ in this frame field by
$\hat{\nabla}_i e_k = \hat{\Gamma}_i\,^j\,_k\,e_k$.
If we express now all tensor fields (except the $e_k$) as well as
equations (\ref{wconnequ}), (\ref{wLequ}), (\ref{dequ}) and the condition
that
$\hat{\nabla}$ be torsion free in terms of the frame field and the
connection coefficients, we get for the unknown
\begin{equation}
\label{gcfeunknown}
u = (e^{\mu}\,_k,\,\,\,\,\,\hat{\Gamma}_i\,^j\,_k,\,\,\,\,\,\hat{L}_{jk},
\,\,\,\,\,d^{i}\,_{jkl}),  
\end{equation}
the system of equations
\begin{equation}
\label{gentor}
[e_{p},e_{q}] =
(\hat{\Gamma}_{p}\,^{l}\,_{q} - \hat{\Gamma}_{q}\,^{l}\,_{p})\,e_{l},
\end{equation}
\begin{equation}
\label{gencorv}
e_{p}(\hat{\Gamma}_{q}\,^{i}\,_{j}) - 
e_{q}(\hat{\Gamma}_{p}\,^{i}\,_{j}) 
- \hat{\Gamma}_{k}\,^{i}\,_{j}(\hat{\Gamma}_{p}\,^{k}\,_{q} -
\hat{\Gamma}_{q}\,^{k}\,_{p}) +
\hat{\Gamma}_{p}\,^{i}\,_{k} \hat{\Gamma}_{q}\,^{k}\,_{j} 
- \hat{\Gamma}_{q}\,^{i}\,_{k} \hat{\Gamma}_{p}\,^{k}\,_{j}  
\end{equation}
\[
= 2\,\{g^{i}\,_{[p}\,\hat{L}_{q]j} 
- g^{i}\,_{j}\,\hat{L}_{[pq]} 
- \,g_{j[p}\,\hat{L}_{q]}\,^i\}
+ \Theta \, d^{i}\,_{jpq}, 
\]
\begin{equation}
\label{gencbian}
\hat{\nabla}_{p}\,\hat{L}_{qj} - \hat{\nabla}_{q}\,\hat{L}_{pj} 
= d_{i} \, d^{i}\,_{jpq},
\end{equation}
\begin{equation}
\label{genbian}
\nabla_{i} d^{i}\,_{jkl} = 0.
\end{equation}
In the last equation the connection $\nabla$ is used. This poses
no problem because the connection coefficients of $\nabla$ in the frame
$e_k$ are related to those of $\hat{\nabla}$ by the purely algebraic
relation $\hat{\Gamma}_i\,^j\,_k = \Gamma_i\,^j\,_k
+ \delta^j\,_i\,f_k + \delta^j\,_k\,f_i - \eta_{ik}\,\eta^{jl}\,f_l$,
whence $f_i = 1/n\,\hat{\Gamma}_i\,^k\,_k$.

No differential equations are given above for the fields $\Theta$ and
$d_k = \Theta\,f_k + \nabla_k \Theta$, they reflect the artificially
introduced \index{conformal!gauge~freedom} conformal gauge
freedom. Note that they cannot be prescribed quite arbitrarily. For a
solution which extends smoothly into a domain containing a set where
$\Theta = 0$, we should have $d = \nabla \Theta$ on this set.

To obtain the metric conformal field equations, (\ref{dequ}),
(\ref{cgequ}), (\ref{Lequ}) have been complemented by equations for
the conformal factor which contain basic information on Einstein's
equations.  Finding a suitable gauge to obtain useful evolution
equations was then considered a different issue. In the case of the
equations above we will combine these two steps into one. The extended
gauge freedom will be used here in a very special way. The question
whether there are other useful possibilities has not been
investigated.

We begin by describing the construction of a gauge. Let $\tilde{S}$ be a
space-like hypersurface in the given vacuum solution 
$(\tilde{M}, \tilde{g})$. We choose on $\tilde{S}$ a positive `conformal
factor' $\Theta_*$, a frame field $e_{k*}$, and a 1-form $b_*$
such that $\tilde{g}(e_{i*}, e_{k*}) = \Theta^2_*\,\eta_{ik}$  and
$e_{0*}$ is orthogonal to $\tilde{S}$. 
Then there exists through each point $x_* \in \tilde{S}$ a unique
conformal geodesic
$x(\tau)$, $b(\tau)$ with $\tau = 0$ on $\tilde{S}$ which satisfies there
the initial conditions $\dot{x} = e_{0*}$, $b = b_*$. These curves define
a smooth caustic free congruence in a neighbourhood $U$ of
$\tilde{S}$ if all data are smooth. Furthermore, $b$ defines a smooth
1-form  on $U$ which supplies a Weyl connection  
$\hat{\nabla}$ as described above. A smooth frame field $e_k$ and
conformal factor $\Theta$ are then obtained on $U$ by solving
(\ref{Wccgxequ}), (\ref{Wdcgxequ}) for given initial conditions $e_{k} =
e_{k*}$, $b = b_*$ on $\tilde{S}$. The frame field is orthonormal
for the metric $g = \Theta^2\,\tilde{g}$, i.e.
$g_{ik} \equiv g(e_i, e_k) = diag(1, -1, -1, -1)$. Dragging along local
coordinates
$x^{\alpha}$, $\alpha = 1, 2, 3$ on $\tilde{S}$ with the congruence and
setting $x^0 = \tau$ we obtain a coordinate system.
In this gauge we have on $U$
\begin{equation}
\label{wgaugecond}
\dot{x} = e_0 = \partial_{\tau},\quad
\hat{\Gamma}_0\,^j\,_k = 0,\quad
\hat{L}_{0k} = 0.
\end{equation}
Coordinates, a frame field, and a conformal gauge as above will be
referred to as a \index{conformal!Gauss~system} {\it conformal Gauss system}.

We could now derive from (\ref{Wdcgxequ}) and the conformal geodesics
equations an equation for 
$d_k = \Theta\,b_k = \Theta\,f_k + \nabla_k \Theta$
and use it together with (\ref{Wdcgxequ}) to complement the system
(\ref{gentor}) to (\ref{genbian}).
Surprisingly, however, the fields $\Theta$ and $d_k$ can be determined 
explicitly  (\cite{friedrich:AdS}): {\it  If $\tilde{g}$ is a
solution to Einstein's vacuum equations with cosmological constant
$\lambda$, the fields 
$\Theta$ and  $d_k$ are given in our gauge by the explicit expressions
\begin{equation}
\label{Thetaexpl}
\Theta = \Theta_*\,\left(1 + \tau\,<b_*, \dot{x}_*>
+ \frac{\tau^2}{2}\,\left(
\Theta_*^{-2}\,\frac{\lambda}{6} + \frac{1}{2}\,
g^{\sharp}(b_*, b_*)\right)\right),
\end{equation}
\begin{equation}
\label{bexpl}
d_0 = \dot{\Theta},\,\,\,\,\,\,d_a = \,<b_*,\Theta_*\,e_{a*}>,
\,\,\,a = 1, 2, 3, 
\end{equation}
where the quantities with a subscript star are considered as constant
along the conformal geodesics and $g^{\sharp}$ denotes the
contravariant version of $g$}. 

With these expressions (\ref{gentor}), (\ref{gencorv}), (\ref{gencbian}),
(\ref{genbian}) provide a complete system for $u$. In spite of the fact
that we use a special gauge, we refer to these equations as the
\index{conformal!field~equations!general} {\it general conformal field
equations} to indicate that they employ the full gauge freedom preserving
a conformal structure.

One might wonder why in the present case it should be sufficient to give
the explicit expression for $\Theta$ and $b$ while in the case of the
metric conformal field equation we had to use Einstein's equations again
to obtain equations for the conformal factor. However, the occurrence of
$\lambda$ shows that information about Einstein's equations
is encoded in (\ref{Thetaexpl}), (\ref{bexpl}). We note the following
important consequence. If we have a sufficiently
smooth solution to the metric conformal field equations in a domain
comprising a non-empty set ${\cal J} = \{\Omega = 0, d\,\Omega \neq 0\}$,
it follows from (\ref{lambdaequ}) that
$\nabla_{\rho}\,\Omega\,\nabla^{\rho}\,\Omega = - \lambda/3$ 
on ${\cal J}$, i.e. the causal nature of ${\cal J}$ is determined by the
field equations (\cite{penrose:scri:let}). The analogous result 
\begin{equation}
\label{scricaus}
\nabla_k\,\Theta\,\nabla^k\,\Theta = - \frac{1}{3}\,\lambda
\quad\mbox{where}\quad \Theta = 0,  
\end{equation}
follows in the present case from (\ref{Thetaexpl}), (\ref{bexpl}) and the
definition of $b$ if the congruence does
not degenerate where $\Theta$ vanishes (\cite{friedrich:AdS}).

Our gauge is not only distinguished by the fact that it is provided by
the conformal structure itself and the explicit information  on $\Theta$
and $b$, but also by a remarkable simplicity of the resulting evolution
equations. Setting
$p = 0$ and observing the gauge conditions 
(\ref{wgaugecond}) we obtain from (\ref{gentor}) - (\ref{genbian})
\[
\partial_{\tau}\,e^{\mu}\,_{q} =
- \hat{\Gamma}_{q}\,^{l}\,_{0}\,e^{\mu}\,_{l},
\]
\[
\partial_{\tau}\,\hat{\Gamma}_{q}\,^{i}\,_{j} = 
- \hat{\Gamma}_{k}\,^{i}\,_{j}\,\hat{\Gamma}_{q}\,^{k}\,_{0} 
+ g^{i}\,_{0}\,\hat{L}_{qj} 
+ g^{i}\,_{j}\,\hat{L}_{0q} 
- \,g_{j0}\,\hat{L}_{q}\,^i
+ \Theta \, d^{i}\,_{j0q}, 
\]
\[
\partial_{\tau}\,\hat{L}_{qj} 
= d_{i} \, d^{i}\,_{j0q},
\]
\[
\nabla_{i} d^{i}\,_{jkl} = 0.
\]
Extracting from the Bianchi equation by the methods discussed above a
symmetric hyperbolic system, we get symmetric hyperbolic reduced
equations for those components of $u$ which are not determined explicitly
by the gauge conditions. 
It has been shown in \cite{friedrich:AdS} for such a choice of reduced
equations (assuming (\ref{Thetaexpl}), (\ref{bexpl})) that any solution
which satisfies the constraints on a suitable space-like hypersurface does
indeed satisfy the complete set of field equations in the part of the
domain of dependence of the initial data set where $\Theta$ is
positive. 

A system of similar simplicity is obtained if we use metric Gauss
coordinates and the vacuum Bianchi identity to solve the vacuum field
equations directly (\cite{friedrich:1981b}). While it is well known,
however, that in general Gauss coordinates tend to develop caustics
quickly, it will be seen below that conformal Gauss coordinates can be
arranged to cover large space-time domains or even entire maximal
globally hyperbolic space-times in non-trivial cases.

Our main purpose in reformulating the field equations was to obtain
results about their solutions. In fact, all the properties above have been
used to establish properties of solutions by formulating and
analyzing boundary value problems which take into account as far as
possible the geometric nature of the situations under study. In the
following we shall indicate the kind of questions which have been asked
and the type of results obtained so far.

\section{The Penrose proposal}

Experience with other theories governed by hyperbolic field equations
suggests that Einstein's theory should admit wave phenomena. For the same
reason for which some work is required to bring to light the hyperbolic
nature of Einstein's equation, we need to make an effort to exhibit the wave
phenomena. Since the theory only supplies the metric and the associated
curvature field, we need to explain which part of the gravitational field
shows wave-like behaviour with respect to which reference structure and to
what extent wave phenomena observed in mathematical expressions are of any
physical significance. The standard textbook discussion of gravitational
radiation bypasses this problem by postulating some ad hoc background
solution and studying solutions to the linearized equations on it. 
This method allows one to apply the machinery developed in other
areas of field theory, but specific features of Einstein's theory arising
from its non-linearity are suppressed. Nowadays there exist much more
sophisticated approximation methods, which also take into account
insights from the approach to be described below (cf. \cite{blanchet},
\cite{thorne83}). 

Work initiated in the late 1950's by Pirani, Trautman, Bondi and others
led to a concept of radiation which involves an `approximation to nature',
namely the idealization of an \index{isolated system} {\it isolated system}, but which
does not require for its definition any mathematical approximations in
the given framework. Clearly, the gravitational field can be expected to
develop a wave-like character only sufficiently far away from the
generating system. The idea then is to consider a gravitating system,
consisting of  stars, black holes, or bounded systems of stars like
galaxies, which is so far away from other such systems that the latter can
essentially be ignored and taken into account only in terms of the
gravitational radiation emitted by them. Thus one is led to study the {\it
far field} of this system.

Following a light ray which escapes from the system to infinity we may
expect the field to become weaker and weaker and in fact {\it 
asymptotically flat} in the sense that the metric field approaches that
of a limit Minkowski space. If some substructure, e.g. some component of
the Weyl curvature, implies a non-trivial field on this limit space it
might attain a natural interpretation as the radiation escaping from the
system. 

To convert this idea into a workable definition two things are needed: a
precise description of the limit procedure and detailed information on the
decay of the metric and the curvature field. Bondi et al.
\cite{bondi:et.al}, Newman and Penrose \cite{newman:penrose}, and Sachs
\cite{sachs:waves} specify the limits in terms of so-called {\it Bondi
coordinates}, which are generated by distinguished families of 
outgoing null hypersurfaces which extend in the future smoothly to infinity.
On the hypersurfaces a convenient parameter is specified on the generating
null geodesics which themselves are parameterized by spherical coordinates.
This choice of coordinates is quite natural for analyzing radiation
phenomena, since for hyperbolic field equations perturbations of the field
essentially travel along characteristics. 

The most delicate part of the whole project is to gain information on the
decay. Obviously, we cannot prepare the systems of interest and let them
go to observe the decay of the field. In fact, at present the main
interest in our whole subject is to obtain information on wave forms
mathematically to possibly help identify gravitational radiation in the
signals supplied by the measuring devices. To prepare the systems and
study their evolution for an infinite time by purely mathematical means
was out of the question at a time when not even the Cauchy problem local
in time was completely understood. Therefore the authors above had to
rely on their experience with a few model examples, linearized theory,
assumptions of simplicity, physical intuition etc. to make a reasonable
guess. 

In \cite{bondi:et.al}, \cite{sachs:waves} the asymptotic behaviour
was imposed by requiring that certain metric coefficients behave in
Bondi coordinates similarly as in the case of Minkowski space while in
\cite{newman:penrose} fall-off conditions for certain components of the
Weyl curvature were specified. In both cases the assumptions proved
consistent with the first few steps of a formal expansions involving the
field equations. The authors could associate certain asymptotic data
convincingly with the gravitational radiation field, and conclude from
the behaviour of the {\it Bondi-energy-momentum} that outgoing
gravitational waves carry only positive energy (\cite{bondi:et.al},
\cite{sachs:waves}).

Soon afterwards a new idea brought into the open the basic geometric
structure underlying this analysis. In \cite{penrose:scri:let},
\cite{penrose:scri} Penrose proposed to {\it characterize the fall-off
behaviour of gravitational fields in terms of the extensibility of the
conformal structure through null infinity}. 

To illustrate the idea we begin by describing certain conformal embeddings
of the simply connected, conformally flat standard solutions to the vacuum
field equations $\mathrm{Ric}[\tilde{g}] = \lambda\,\tilde{g}$ into the 
\index{Einstein!cosmos} {\it Einstein cosmos} (cf. also \cite{hawking:ellis}). The latter
is given by the manifold $\bar{M} = \mathbb{R} \times S^3$, endowed with
the product line element $g = d\,s^2 - d\,\omega^2$ where $d\,\omega^2 =
d\,\chi^2 +
\sin^2\chi\,d\,\sigma^2$ denotes the standard line element on the
3-sphere with
$d\,\sigma^2 = d\,\theta^2 + \sin^2 \theta\,d\,\phi^2$ the  
standard line element on the 2-sphere.
We shall in the following be mainly interested in the conformal extension of
Minkowski space and shall use the other two examples only to illustrate
certain arguments. 

The case $\lambda = 0$: We express the metric of {\it Minkowski space}
$(\tilde{M} \simeq  \mathbb{R}^4, \tilde{g}  =
\eta_{\mu \nu}\,d\,x^{\mu}\,d\,x^{\nu})$ in spherical
coordinates such that
$\tilde{g} = dt^2 - (dr^2 + r^2\,d\,\sigma^2)$ and define
a diffeomorphism $\Phi$ of $\tilde{M}$ onto the open subset 
$\Phi(\tilde{M}) = \tilde{M}' \equiv 
\{|s + \chi| < \pi,\,\,\,\,|s - \chi| < \pi,\,\,\,\, \chi
\ge 0
\}$ of $\bar{M}$ such that $\Phi^{-1}$ is given by
\[
t = \frac{\sin s}{\cos s + \cos \chi},\,\,\,\,\,\,\,\,\,
r = \frac{\sin \chi}{\cos s + \cos \chi}.
\]
It follows that $\Omega^2\,\Phi^{-1*}\tilde{g} = g$ on 
$\tilde{M}'$ with the conformal factor 
$\Omega = \cos\,s + \cos\,\chi$ which is positive on $\tilde{M}'$.
The embedded Minkowski space acquires then a {\it conformal boundary}
${\cal J}$ in $\bar{M}$, which is defined as the subset
${\cal J} \equiv \partial\,\tilde{M}' = 
{\cal J}^- \cup {\cal J}^+ \cup \{i^-, i^0, i^+\}$,
of $M$, where the hypersurfaces 
${\cal J}^{\pm} = \{ \tau \pm \chi = \pm \pi,\, {0} < \chi < \pi\}$, which
can be thought of as being generated by endpoints of null geodesics,
represent {\it future} resp. {\it past null infinity}, the points
$i^{\pm} = \{ \chi = 0, \tau = \pm \pi \}$ represent 
{\it future} resp. {\it past time-like infinity} and the point 
$i^0 = \{ \chi = \pi, \tau = 0 \}$ represents {\it space-like infinity}.
The key point here is that the embedded Minkowski space together with its
conformal structure and the conformal factor $\Omega$ can be extended
smoothly to the manifold with boundary
$M = \tilde{M}' \cup {\cal J}^+ \cup {\cal J}^-$ such that 
$\Omega = 0, d\,\Omega \neq 0$ on the null hypersurfaces 
${\cal J}^{\pm} \simeq  \mathbb{R} \times S^2$.

It may be of interest to note that H. Weyl already in 1923 considers the
conformal embedding of Minkowski space into a quadric in the five
dimensional real projective space in a relativity inspired discussion of the
M\"obius group and refers to the set which corresponds to
${\cal J}$ as to the `sphere at infinity' (\cite{weyl}). The fact that
Weyl did not see, in spite of its beautifully geometric nature, the
generalization proposed in \cite{penrose:scri:let}, indicates to what 
extent the kind of questions being asked and, in fact, our whole view of
the subject have changed since then. 

The case $\lambda < 0$: The {\it de Sitter space} is given by 
$(\tilde{M} =  \mathbb{R} \times S^3, \tilde{g} 
= dt^2 - \cosh^2t\,d\,\omega^2)$. The map
$\tilde{M} \ni (t, \,\,\vartheta) \stackrel{\Phi}{\rightarrow} 
(s = \arctan e^t - \frac{\pi}{4}, \,\,\vartheta) \in \bar{M}$, where
$\vartheta$ denotes the points in $S^3$, maps de Sitter space
diffeomorphically onto the open subset $\tilde{M}' =  [- \pi/4, \pi/4]
\times S^3$ of the Einstein cosmos. It satisfies
$\Omega^2\,\Phi^{-1*}\tilde{g} = g$ on $\tilde{M}'$ with the positive
conformal factor $\Omega = \cos\,(2\,s)$. Again the embedded space
together with its conformal structure and the conformal factor $\Omega$ can
be extended smoothly to a manifold with boundary
$M = \tilde{M}' \cup {\cal J}^+ \cup {\cal J}^-$ such that 
$\Omega = 0, d\,\Omega \neq 0$ on the space-like hypersurfaces 
${\cal J}^{\pm} = \{\pm \pi/4\} \times S^3$ of $M$.

The case $\lambda > 0$. In spherical coordinates we obtain {\it anti-de
Sitter covering space}  in the form $(\tilde{M} =  \mathbb{R}^4,
\tilde{g} = \cosh^2r \,dt^2 - dr^2 - \sinh^2r\,d\sigma^2)$. The map
$\tilde{M} \ni (t, r, \theta, \phi) \stackrel{\Phi}{\rightarrow}
(s = t, \chi = 2\,\arctan\,(e^r) - \frac{\pi}{2}, \theta, \phi) \in \bar{M}$
defines a diffeomorphism of $\tilde{M}$ onto the subset
$\tilde{M}' = \{\chi < \pi/2\}$ of the Einstein cosmos which satisfies
$\Omega^2\,\Phi^{-1*}\tilde{g} = g$ on 
$\tilde{M}'$ with the positive conformal factor $\Omega = \cos\chi$.
The embedded space together with its conformal structure and the conformal
factor $\Omega$ can be extended smoothly to a manifold with boundary
$M = \tilde{M}' \cup {\cal J}$ such that 
$\Omega = 0, d\,\Omega \neq 0$ on the time-like hypersurface 
${\cal J} = \{\chi = \pi/2\} \simeq  \mathbb{R} \times S^2$ of $M$.

The conformal extensions above are of interest already because they provide
detailed information about the global causal relations in the
three space-times and allow us to discuss the asymptotic behaviour of
solutions to conformally invariant  equations such as the Maxwell
equations easily and with any desired precision. For our present
discussion they are of particular interest because the three
conformally embedded spaces and the given conformal factors provide {\it
solutions of the metric conformal field equations which extend, as
solutions, through the conformal boundaries}. The question whether this
fact can be considered as a model for the behaviour of more general
gravitational fields leads us (in hindsight) to the generalization
proposed in \cite{penrose:scri:let}. 

The embedding formulae given above are distinguished by their simplicity
but may look otherwise somewhat arbitrary. It turns out that we do
not have much freedom. For given $\vartheta \in S^3$ the curve 
$ \mathbb{R} \ni \tau
\rightarrow  (s = 2\,\arctan \frac{\tau}{2}, \,\,\vartheta) \in \bar{M}$ is a
conformal geodesic on the Einstein cosmos. If $\chi(\vartheta) \neq 0, \pi$,
this curve intersects the conformal boundary ${\cal J}^{\pm}$ of the
embedded Minkowski space for values $\tau_{\pm}(\vartheta) \neq 0$ with
$\tau_- = - \tau_+$. By specifying suitable initial data on the
hypersurface $\{t = 0\}$, this curve can be constructed completely in
terms of Minkowski space and we find that the conformal factor
$\Theta$ obtained by solving (\ref{Wdcgxequ}) along this conformal
geodesic remains positive on Minkowski space and vanishes precisely as
$\tau
\rightarrow \tau_{\pm}$, where $\dot{\Theta}(\tau_{\pm}) \neq 0$. 
It follows that, apart from a coordinate transformation and a rescaling with
a conformal factor $\omega$ which is positive on $\bar{M}$, 
the conformal extension can be constructed completely by integrating a
suitable congruence of conformal geodesics and an associated conformal
factor on Minkowski space and extending the resulting formulae by continuity
(cf. \cite{friedrich:cg on vac} for details).

We could have surmised the possibility for this already when we
encountered formula (\ref{Thetaexpl}). It shows that for suitable
initial data $\Theta$ has precisely two zeros. This also raises a
question. Does the fact that the formula is valid on a general vacuum
solutions suggest that a similar procedure works under much more
general assumptions? We are thus led again to the generalization
proposed in \cite{penrose:scri:let}.

Stating in general terms what has been found above, one arrives at the
following definition. 

\begin{definition}
\label{assidef}
A smooth space-time $(\tilde{M}, \tilde{g})$ is called 
\index{asymptotically simple} {\rm asymptotically simple} if there exists a smooth, oriented,
time-oriented, causal space-time
$(M, g)$ and on $M$ a smooth function $\Omega$ such that:\\ 
(i) $M$ is a manifold with boundary ${\cal J}$,\\
(ii) $\Omega > 0$ on $M \setminus {\cal J}$ and $\Omega = 0$, $d\,\Omega
\neq 0$ on ${\cal J}$,\\
(iii) there exists an embedding $\Phi$ of $\tilde{M}$ onto 
$\Phi(\tilde{M}) = M \setminus {\cal J}$
such that $\Omega^2\,\Phi^{-1*}\tilde{g} = g$.\\  
(iv) each null geodesic of $(\tilde{M}, \tilde{g})$ acquires two distinct
endpoints on ${\cal J}$. 
\end{definition}

This definition, which only involves the conformal structure of $(M, g)$,
reflects the differential geometric aspects of the situations considered
above. It implies restrictions on the global
structure of the space-time. Condition $(iv)$, for instance, implies that all
null geodesics are complete. The set ${\cal J}$ thus represents a boundary
at null infinity which can be thought of as being generated by ideal
endpoints of null geodesics. Conditions $(i)$, $(ii)$ are somewhat
redundant, since $(ii)$ implies already that ${\cal J}$ is a smooth
hypersurface. However, this is not the only purpose of $(ii)$. Together
with $(iii)$ it specifies how precisely
$\Phi^{-1*}\tilde{g}$ is to be rescaled to obtain a smooth, non-degenerate
metric.  The definition above was motivated by the following new idea 
(\cite{penrose:scri:let}). 

\index{Penrose proposal} {\bf Penrose proposal}: {\it Far fields
of isolated systems behave like asymptotically simple space-times in the
sense that they can be smoothly extended to null infinity, as indicated
above, after suitable conformal rescalings}. 

It should be noted that we have only tried to indicate the idea here. It
should be applied with understanding and the definition above may
require modifications in certain situations. For instance, in applying the
definition to physically realistic situations it may be advisable to
replace $(iv)$ by some other completeness conditions (cf.
\cite{geroch:horowitz} for an example) to allow for the possibility that
some of the null geodesics enter black holes. The standard example here is
the Schwarzschild-Kruskal solution, which has two asymptotic ends which
admit smooth conformal extensions (cf. B. Schmidt, this volume). We
consider any such changes as minor adaptations. As we shall see below,
there may be required much more subtle ones. 

While in the earlier description of fall-off conditions coordinates play
a crucial role and while even in case of asymptotically simple
space-times it is often convenient to use Bondi coordinates, the present
characterization is completely independent of any distinguished
coordinate systems and therefore much more flexible than the earlier
ones. But the beauty of the proposal is that it brings out clearly the
key geometrical structure. 

Central and critical is the requirement $(iii)$ and the smoothness
assumption. The smoothness properties of the rescaled field $g$ are an
expression for the fall-off behaviour of the physical field
$\tilde{g}$. In view of the Penrose proposal definition (\ref{assidef})
acquires suddenly a much stronger interest. It is now in competition with
another requirement and involves an extremely sharp characterization of
the decay of fields which are governed by the quasi-linear,
gauge hyperbolic field equations.  

The question whether the assumption that the field equations hold near 
${\cal J}$ is consistent with the smoothness and conditions $(i)$ --
$(iii)$, or whether the smoothness assumptions made in definition
(\ref{assidef}) need to be relaxed has been and, as seen below, still is
a matter of debate.  The question is not whether $C^{\infty}$ should be
replaced by 
$C^k$ for some large $k$. The question is whether solutions to the field
equations admit conformal extensions of class $C^k$ where $k$ can be
chosen large enough to make the proposal meaningful and, related to this,
whether the conformal Weyl tensor tends to zero on the conformal boundary.

Penrose addresses this problem in \cite{penrose:scri}. For later
comparison we recall the assumptions and give (a version of) the main
argument, using the spin frame formalism. Consider a smooth solution to
the metric conformal field equations with $\lambda = 0$ on the physical
manifold
$\tilde{M}$ and {\it assume that it admits a conformal extension to a
manifold} $M = \tilde{M} \cup {\cal J}^+$ {\it of class} $C^4$ 
{\it such that} $g$ {\it and} $\Omega$ {\it extend to fields in}
$C^3(M)$. The conformal Weyl spinor $\Psi_{abcd}$ is then in $C^1(M)$.

By definition, $\Omega = 0$, $d\,\Omega \neq 0$ and, as
observed earlier, $\nabla_{aa'}\Omega\,\nabla^{aa'}\Omega = 0$ on 
${\cal J}^+$. We assume that there is a space-like surface $\Sigma$ in
${\cal J}^+$ which intersects each null generator of ${\cal J}^+$
exactly once. In a neighbourhood of $\Sigma$ there are then $C^3$ spinor
fields $o^a$, $\iota^a$ with $\iota^a\,o_a = 1$ such that
$\nabla^{aa'}\Omega = - o^a \bar{o}^{a'}$ on ${\cal J}^+$ and the
complex vector field $m^{aa'} = o^a \bar{\iota}^{a'}$ is tangent to
$\Sigma$. Note that the set of such spin frames defines a reduction 
$U(\Sigma)$ of the bundle of normalized spin bases over $\Sigma$ to the
subgroup $U(1)$ of $SL(2, \mathbb{C})$, 
where $e^{i\frac{\theta}{2}} \in U(1)$ acts on the spin frames by 
$(o, \iota) \rightarrow (e^{i\frac{\theta}{2}}\,o,
e^{-i\frac{\theta}{2}}\iota)$. 
A $2:1$ bundle morphism maps $U(\Sigma)$ onto the bundle
$O^+(\Sigma)$ of oriented frames which are orthonormal for the induced
metric $h$ on $\Sigma$ such that we can write 
$m = 1/\sqrt{2}\,\,(r_1 + i\,r_2)$ with $r \equiv \{r_A\}_{A = 1, 2} \in
O^+(\Sigma)$. Denote by $u$ the generator of the structure group $SO(2)$
of $O^+(\Sigma)$ with components 
$u^A\,_B = \delta^A\,_1\,\delta^2\,_B - \delta^A\,_2\,\delta^1\,_B$ and
by $Z$ the vertical vector field on $O^+(\Sigma)$ generated by $u$.

Equation (\ref{Omegaequ}) reduces on ${\cal J}^+$ to 
$\nabla_{aa'}\,\nabla_{bb'}\,\Omega = s\,\epsilon_{ab}\,\epsilon_{a'b'}$.
Since $o^a\nabla_{aa'}$ is an interior operator on ${\cal J}^+$ it
follows that  
$o^a\nabla_{aa'}(o_b \bar{o}_{b'}) = - s\,o_b\,\epsilon_{a'b'}$  on
${\cal J}^+$ from which we obtain by suitable contractions 
$\iota^a\nabla_m\,o_a + \bar{\iota}^{a'}\nabla_m\,\bar{o}_{a'}$ whence 
$\gamma \equiv g(\nabla_m m, \bar{m}) = - 2\,\iota^a\nabla_m\,o_a$. 
Given $\gamma$, the connection ${\cal D}$ of $h$ is known, because
${\cal D}_m\,m = - \gamma\,m$, ${\cal D}_m\,\bar{m} = \gamma\,\bar{m}$.

With $\phi_{abcd} = \Omega^{-1}\Psi_{abcd}$ the Bianchi equation
(\ref{scvBian}) takes on $\tilde{M}$ the form
\begin{equation}
\label{bianrewr}
\nabla^a\,_{a'}\,\Omega\,\Psi_{abcd}
- \Omega\,\nabla^a\,_{a'}\,\Psi_{abcd} = 0.
\end{equation}
(We note that in the case where $0 \neq \lambda 
= - 3\,\nabla_{ff'}\Omega\,\nabla^{ff'}\Omega$ on ${\cal J}$ this
implies at each point of ${\cal J}$ immediately that
$0 = - 3\,\nabla_{ea'}\Omega\,\nabla^{aa'}\Omega\,\Psi_{abcd} = 
\lambda\,\,\Psi_{ebcd}$.) It follows that   
$\Psi_{abcd} = \alpha\,o_a\,o_b\,o_c\,o_d$ on ${\cal J}^+$ with
$\alpha = \Psi_{abcd}\,\iota^a\,\iota^b\,\iota^c\,\iota^d$. 
{\it If in addition it is assumed that}
\begin{equation}
\label{Pcond}
\Omega\,\nabla_{ee'}\,\nabla^a\,_{a'}\,\Psi_{abcd}
\rightarrow 0 \quad\mbox{at}\quad {\cal J}^+,
\end{equation}
then equation (\ref{bianrewr}) implies, after taking a covariant
derivative, symmetrizing in $a'$ and $e'$, and performing some spinor
algebra,  the equation $o^e\,\nabla_{ee'}\,\Psi_{abcd} = 0$ on
${\cal J}^+$. With the special form of the conformal Weyl spinor on
${\cal J}^+$ it follows after contraction with $\bar{\iota}^{b'}$
\begin{equation}
\label{pequ}
\nabla_m\,\alpha  - 2\,\gamma\,\alpha = 0
\quad\mbox{on}\quad {\cal J}^+.
\end{equation}  
By its definition the $C^1$ function $\alpha$ can be considered as a
$0$-form on $O^+(\Sigma)$ of spin weight $- 2$, i.e.
\begin{equation}
\label{spwalpha}
Z\,\alpha = - 2\,i\,\alpha \quad\mbox{on}\quad O^+(\Sigma).
\end{equation}
Equation (\ref{pequ}) says that the covariant
differential of $\alpha$ on $O^+(\Sigma)$ (cf. \cite{dieudonne:IV})
vanishes on vector fields which project onto $m$. It can thus be expressed
in the form
\begin{equation}
\label{alphapequ}
(H_1 + i\,H_2)\,\alpha = 0
\quad\mbox{on}\quad O^+(\Sigma),
\end{equation}
where for $A = 1, 2$ we denote by $H_A$ the horizontal vector
field on $O^+(\Sigma)$ which projects at the point $r \in
O^+(\Sigma)$ onto the tangent vector $r_A$ of $\Sigma$.

{\it Suppose now that $\Sigma$ is diffeomorphic to the two-sphere $S^2$}. 
Because $S^2$ carries only one conformal structure the conformal scaling
of $g$ can be chosen near $\Sigma$ such that $(\Sigma, h)$ can be
identified isometrically with $(S^2, - d\,\sigma^2)$. This allows us to
identify $O^+(\Sigma)$ with $SO(3)$, the isometry group of $S^2$, such
that there are left invariant vector fields $Y_i$, $i = 1,2,3$, on 
$SO(3)$ which satisfy under this identification $Y_1 = H_2$, $Y_2 = H_1$,
$Y_3 = Z$ and $[Y_i, Y_j] = \epsilon_{ijk}\,Y_k$, where 
$\epsilon_{ijk}$ is totally anti-symmetric with $\epsilon_{123} = 1$.
Equations (\ref{spwalpha}), (\ref{alphapequ}) then imply the
distributional equation 
\[
- \frac{1}{2}(Y_1^2 + Y_2^2 + Y_3^2)\,\alpha = \alpha
\quad\mbox{on}\quad O^+(\Sigma). 
\]
Since the operator on the left hand side is the Casimir operator with
respect to the Killing form on $SO(3)$, the solution space of this
equation is well known (cf. \cite{sugiura}). It is spanned by the
matrix elements of an irreducible three-dimensional unitary
representation  of the group $SU(2)$ (in the notation of
\cite{friedrich:i-null} the functions $T_2\,^i\,_j$), which descends to
$SO(3)$. Because the only function in this
space which satisfies (\ref{spwalpha}) is $\alpha = 0$, it follows that
the Weyl spinor field vanishes on $\Sigma$. Thus, $\phi_{abcd}$
extends in a continuous fashion to ${\cal J}^+$ and, as a consequence,
satisfies Sachs' {\it peeling property} (\cite{penrose:scri},
\cite{sachs:bdrycond}, \cite{sachs:waves}).

Remarkable as it is that such a conclusion can be drawn from the
special conformal covariance of the Bianchi equations, the fact that
we are dealing with a spin-2 field, and the topological assumption on
$\Sigma$, there remains the question whether it
follows from the nature of the long time evolution by the field
equations that the smoothness assumptions made above and (\ref{Pcond})
can be satisfied and whether this is in fact the case for a sufficiently
large class of solutions. 

There are two different issues here. We start with a discussion of the
situation where the vacuum field equations with a cosmological
constant $\lambda$ are satisfied {\it locally near a point $p$ of
${\cal J}$} and where $\Omega$ and $g$ are smooth. As seen earlier, the
conformal field equations then imply
$\nabla_{\rho}\,\Omega\,\nabla^{\rho}\,\Omega = - \lambda/3$. In
particular, ${\cal J}$ is a null hypersurface if $\lambda = 0$. In that
case we can expect ${\cal J}$ to have two components ${\cal J}^{\pm}$
and $p \in {\cal J}^+$. To get some idea of the class of solutions
which are compatible with conditions $(i)$ -- $(iii)$, we consider the
{\it asymptotic characteristic initial value problem} 
\index{initial~value~problem!characteristic!asymptotic}for the metric conformal field
equations where data are prescribed on two intersecting hypersurfaces
which are to assume, by the arrangement of the initial data on them,
the meaning of an outgoing null hypersurface ${\cal N}$ which meets
${\cal J}^+$ in a space-like 2-surface $\Sigma$ and the part
${\cal J}^{+}_*$ of ${\cal J}^+$ in the past of $\Sigma$. This is
essentially the problem underlying the formal expansions studied in
\cite{bondi:et.al}, \cite{newman:penrose}, \cite{sachs:waves}.

All solutions for which the rescaled conformal Weyl
tensor $d^i\,_{jkl}$ has a smooth limit on ${\cal J}^{+}_*$ can be
characterized in the analytical as well as in the $C^{\infty}$ case
(\cite{friedrich:1982}, \cite{kannar:char}) in the sense that {\it for
suitably given data the existence of a unique local solution near $\Sigma$
can be shown}. The freedom to prescribe data on
${\cal N}$ and ${\cal J}^{+}_*$, which are in each case two components of
$d^i\,_{jkl}$ referred to as the \index{null!data} {\it null data}, turns out to be
the same as in the standard characteristic initial value problem for
Einstein's vacuum field equations (corresponding to the special geometric
nature of the hypersurface ${\cal J}^{+}_*$, there is a slight difference
with the latter problem concerning the data which may be prescribed in
addition to the null data on the intersection of the two null
hypersurfaces). If, as above, the tensor $d^i\,_{jkl}$ is represented by
the symmetric spinor
$\phi_{abcd}$, then 
$\phi_{abcd}\,\iota^a\,\iota^b\,\iota^c\,\iota^d$ and 
$\phi_{abcd}\,o^a\,o^b\,o^c\,o^d$ denote the null data on ${\cal N}$ and
${\cal J}^{+}$ respectively. It is quite a remarkable property of the
conformal field equations that they define the conformal boundary 
$\{\Omega = 0\}$ in terms of one of their unknowns without restricting the
freedom to specify smooth data there. 

The null data $\phi_{abcd}\,o^a\,o^b\,o^c\,o^d$ on ${\cal J}^{+}$ have
a natural interpretation as the \index{radiation!field!outgoing} {\it
outgoing radiation field}.  Further important physical concepts can be
associated with the hypersurface ${\cal J}^{+}$ or subsets of it and
the questions of interpretation have been extensively analyzed
(cf. also \cite{ashtekar}, \cite{geroch}). All indications suggest
that we are on the right track.

The situation is similar in the de Sitter-type case $\lambda <
0$. Here null infinity is space-like.  We ask whether we can
characterize the solutions which admit a smooth conformal extension at
past null infinity ${\cal J}^-$ under the (natural) assumption that
this hypersurface is orientable and compact as in the case of de
Sitter space and the further assumption that $d^i\,_{jkl}$ extends
smoothly to ${\cal J}^-$.  Again, {\it for suitably given data on
${\cal J}^-$ there can be shown the existence of a unique solution to
the conformal field equations}. The freedom to specify data on
${\cal J}^-$ is essentially the same as in a standard Cauchy problem
with the exception that the mean intrinsic curvature 
of the initial hypersurface cannot be prescribed
(\cite{friedrich:1deSitter}). This appears natural, because
${\cal J}^-$ is a very particular hypersurface with respect to the
solution. Another peculiar feature consists in the fact that the
Hamiltonian constraint becomes trivial on ${\cal J}^-$.

The situation where $\lambda > 0$, in which case null infinity is
time-like, differs in various aspects from the other cases. In this case
the assumption of the smoothness of this limit has a different status
than in the other cases. It appears that in general there would not even
exist complete null geodesics if the field $d^i\,_{jkl}$ tended to blow
up sufficiently fast `far out'. However, {\it there is available a
detailed discussion of the existence of solutions for which} $d^i\,_{jkl}$
{\it has a smooth limit to} ${\cal J}$ (\cite{friedrich:AdS}). 

The solutions studied in these three situations are semi-global in the
sense that in terms of the physical metric their domains are of infinite
extent and even comprise asymptotic regions. However, this is not
sufficient to answer the question posed above. Here comes in the
second issue concerning the compatibility of asymptotic simplicity with
the field equations. {\it What we need to control is the behaviour of the
solutions as they evolve towards null infinity}. Thus, we cannot assume
anything about the field $d^i\,_{jkl}$ (except in the anti-de Sitter-type
solutions where boundary data must be given at space-like and null
infinity to characterize the solutions uniquely) and it is not clear a
priori whether it remains bounded near null infinity. We need to find out
whether it behaves, together with the other unknowns in the conformal
field equations, such that a conformal extension $(M, g)$ of sufficient
smoothness {\it can be constructed}.

There exists a model case which illustrates the type of result we would like
to have. In the case of the de Sitter space considered above ($\lambda = -
3$) we consider the hypersurface $S = \{t = 0\}$. The Cauchy data induced by
de Sitter space on $S$ are given by the standard metric on $S^3$ and a
vanishing extrinsic curvature. It turns out that {\it data for the
equation} $\mathrm{Ric}[\tilde{g}] = \lambda \,\tilde{g}$ {\it which are
sufficiently close to the de Sitter data evolve into solutions which are
asymptotically simple in the sense of definition} (\ref{assidef}), or, in
other words, {\it asymptotic simplicity is non-linearly stable}
(\cite{friedrich:n-geod}, \cite{friedrich:global}). Here {\it
$d^i\,_{jkl}$ attains a smooth limit at the conformal boundary as a
consequence of the evolution process and the assumptions on the initial
data}.  The fact that the Schwarzschild-de Sitter solution has time
slices diffeomorphic to $S^2 \times S^1$ but develops horizons,
singularities, and only patches of a smooth boundary shows 
that neither the compactness of the initial hypersurface nor the sign of the
cosmological constant is sufficient for this behaviour.
Remarkably, it has been shown recently that a solution with a compact
Cauchy hypersurface can admit smooth conformal boundaries in the past as
well as in the future only if the Cauchy hypersurface satisfies the
topological restriction that its fundamental group be finite 
(\cite{andersson:galloway}, cf. G. Galloway, this volume, for further
results concerning de Sitter-type solutions).  
  
An analogous stability result in the Minkowski-type case $\lambda = 0$ is
still lacking. We shall see that a result of this simplicity cannot be
expected.

\section{Asymptotic behaviour of vacuum fields with vanishing
cosmological constant}

For physical applications one would like to know how the sources
interact with each other and with the gravitational field and to what
extent properties of the far field reflect this interaction. For a first
mathematical analysis of the general behaviour of the far fields details
of the sources are irrelevant and it is natural to ignore them in a first
study. Thus we are led to ask whether there exist other solutions to the
Einstein equations $\mathrm{Ric}[\tilde{g}] = 0$ than Minkowski space which
satisfy the requirements of definition (\ref{assidef}).

Two different types of initial value problems have been studied in this
context. The most important one is the 
\index{Cauchy problem!standard} {\it standard Cauchy problem} where
Cauchy data are prescribed on a space-like hypersurface $S$ with one or
several asymptotically Euclidean ends. If there are several such ends we
can expect black holes to develop and condition $(iv)$ to be violated.
Therefore, and because it is the evolution near an end we want to
focus on, we will concentrate on the case $S \simeq \mathbb{R}^3$ with
one end. Standard examples are the hypersurface $\{t = 0\}$ in Minkowski
space with the induced data or perturbations thereof. 

The other type of problem is the 
\index{initial~value~problem!hyperboloidal} {\it hyperboloidal initial value problem}
where data are prescribed on a hypersurface $H$ with boundary $\Sigma$,
which is thought of as a hypersurface in an asymptotically simple
space-time which extends to future null infinity ${\cal J}^+$,
intersects ${\cal J}^+$ in $\Sigma$, and is space-like
everywhere. The boundary $\Sigma$ can have several components, examples
exist in the Schwarzschild-Kruskal space-time (cf. B. Schmidt, this
volume). Again we will concentrate on the case where only one asymptotic
region exists  such that $H \simeq \{x \in \mathbb{R}^3|\,\,|x| \le 1\}$
and $\Sigma \simeq S^2$.  The standard example of such a
hypersurface is given by the extension of the space-like hyperboloid 
$\{ t^2 - |x|^2 = 1, t > 0 \}$ in Minkowski space to ${\cal J}^+$.
Its interior inherits from Minkowski space a metric of constant negative
curvature which can be rescaled to become Euclidean.

While the hyperboloidal initial value problem has been introduced
initially as an intermediate step towards a treatment of the standard
Cauchy problem (\cite{friedrich:hypivp}), it has gained in recent
years an interest in its own as a basis for numerical calculations
(cf. J. Frauendiener and S.  Husa, this volume). It is intrinsically
non-time-symmetric. The future evolution will extend under suitable
assumptions to null infinity while the past time evolution develops a
Cauchy horizon which corresponds to a part $N$ of an outgoing wave
front. The domain covered by its evolution coincides with that which
can be determined by the characteristic initial value problem where
data are prescribed on $N$ (if this set is sufficiently smooth).

\subsection{The hyperboloidal initial value problem}

Since, for reasons which will become clear later, the hyperboloidal initial
value problem is much easier to analyze in the context of asymptotic
simplicity than the standard Cauchy problem, it has been analyzed much
earlier and much more is known about it. If one wants to construct its
solutions by solving the conformal field equations, the first task is to
provide conformal initial data. In the analyses available so far, in which
a non-vanishing constant mean extrinsic curvature is assumed, certain data
can be prescribed on $H$ arbitrarily, certain data are obtained by
solving elliptic equations which degenerate on $\Sigma$, and the remaining
data are obtained by differentiation and algebraic operations which
involve divisions by the conformal factor, which vanishes on $\Sigma$. We
give an outline of the results needed in the following discussion (for 
further information cf. L. Andersson, this volume).

The subtleties arise at $\Sigma$. In \cite{andersson:chrusciel:as},
\cite{friedrich:ACF} free data are prescribed which extend smoothly to
$\Sigma$. It turns out that (a) in general the remaining data have
expansions in terms of powers of $r$ and $\log r$, where $r$ denotes the
conformal distance from $\Sigma$ (i.e. they admit {\it polylogarithmic
expansions} at $\Sigma$), while (b) all data will be smooth if the free
data satisfy certain {\it regularity conditions} which require that a few
functions derived from the free data vanish on $\Sigma$. A much more
general class of data is considered in \cite{andersson:chrusciel:ph}. In
particular, (c) free data are prescribed which admit polylogarithmic
expansions at $\Sigma$ and it is shown that the remaining data also admit
such expansions.

The conformal field equations have been used to study the evolution of some
of these data. {\it In the case} (b) {\it the data evolve into solutions
which satisfy the first three conditions of definition} (\ref{assidef})
{\it and which admit in fact a conformally regular point} $i^+$ {\it at
time-like infinity, analogous to the point} $i^+$ {\it in the case of
Minkowski space, if the data are sufficiently close to Minkowskian
hyperboloidal data} (\cite{friedrich:hypivp}), \cite{friedrich:n-geod},
\cite{friedrich:global}). Recently also {\it the existence of solutions
to the conformal field equations for data of the type} (a) {\it has been
shown on a manifold which contains a boundary hypersurface}
${\cal J}'$ {\it on which the unknown} $\Omega$ {\it of the conformal
field equations vanishes}. The solution is `rough' at ${\cal J}'$ and it
is expected that it admits a polyhomogeneous expansion there 
(\cite{lengard}, cf. also \cite{chrusciel:lengard}). The evolution of
data of type (c) has not been studied yet. 

While the basic idea underlying the Penrose proposal, namely to
characterize the asymptotic behaviour of solutions in terms of conformal
rescalings, is still being used in this study of asymptotically rough
data by employing the conformal field equations, this situation certainly
forbids any straightforward application of definition (\ref{assidef}).
The result seems to indicate that the smoothness assumption in  definition
(\ref{assidef}) may be too strong. However, before we can arrive at a
conclusion here we need to answer the following question: of which physical
relevance are data of type (a) and (c), are they needed to model the systems
of interest and which are the `systems of interest'? 

A clear answer could be given, if it could be shown that standard Cauchy
data by necessity develop into solutions containing hyperboloidal data of
type (a) or (c).  The results for data of type (b) tell us that the decision
about the smoothness of the conformal  structure near null infinity is
essentially made near space-like infinity. Thus we need to analyze the
behaviour of the solution to the standard Cauchy problem there. 

\subsection{On the existence of asymptotically simple vacuum solutions}

We shall see below that the general analysis of the conformal structure of
solutions to the Einstein equations near space-like infinity is quite
complicated. On the other hand, static or, more generally, stationary
asymptotically flat vacuum solutions are known to admit smooth conformal
extensions through null infinity (cf. \cite{dain}).
Thus, to avoid the difficulties discussed below one could try to
construct asymptotically flat standard Cauchy data (on $\mathbb{R}^3$
say) which agree with stationary data near space-like infinity. The time
evolution of these data would then be stationary near space-like infinity
and clearly contain smooth hyperboloidal hypersurfaces. One could then
hope to combine this with the results of \cite{friedrich:n-geod} to show
the existence of solutions to the Einstein equations which satisfy  all
conditions of definition (\ref{assidef}). 

The first attempt at this was made by Cutler and Wald who succeeded in
constructing smooth initial data for the Einstein-Maxwell equations with
arbitrarily small ADM mass which agree with Schwarzschild data near space-like
infinity (\cite{cutler:wald}). Using the results of
\cite{friedrich:global}, they were able to show for the first time the
existence of smooth solutions to the Einstein-Maxwell equations which
satisfy the conditions of asymptotic simplicity. The vacuum case
remained difficult. 

More recently Corvino developed a gluing construction by which an arbitrary
smooth, time-symmetric, asymptotically flat solution $(\mathbb{R}^3, h)$
to the constraints can be deformed on a certain annulus 
$\{x \in \mathbb{R}^3| 0 < R < |x| < 2\,R\}$, such that it extends with a
given finite differentiability to Schwarzschild data outside
$2\,R$ while remaining unchanged inside the radius $R$ and still satisfying
the constraints everywhere  (\cite{corvino}). It turned out difficult,
however, to keep the radius $R$ fixed or to control its growth sufficiently
if the construction is applied to families of initial data for which the
ADM mass $m$ is going to zero. This makes it impossible to obtain the
desired result by an immediate application of \cite{friedrich:n-geod}. 

Quite recently Chru\'sciel and Delay obtained a similar result based on a
modification of Corvino's technique {\it in which the radius $R$ as
well as the differentiability class $C^k$ may be kept fixed while $m
\rightarrow 0$ if they base their construction on initial data 
$(\mathbb{R}^3, h)$ such that in the standard coordinates on
$\mathbb{R}^3$ the parity condition $h(x) = h(-x)$ is satisfied by the
3-metric} (\cite{chrusciel:delay}). In this case it is easy to construct
in a systematic way hyperboloidal hypersurfaces in the time evolution of
the data which contain the part $\{x \in \mathbb{R}^3|\, |x| < 2\,R\}$
of the initial Cauchy hypersurface as a subset and which extend into to
future of the initial hypersurface in the Schwarzschild part of the
evolution. It can be arranged that the induced data approach Minkowskian
hyperboloidal data if $h$ approaches the Euclidean metric on
$\mathbb{R}^3$ (cf. \cite{friedrich:cg on vac} for a construction based on
conformal Gauss systems and \cite{friedrich:i-null} for the discussion of
a smooth limit of Schwarzschild solutions to the Minkowski solution near
space-like infinity). Since this can be done with sufficient
smoothness, the results of \cite{friedrich:n-geod} and the fact that the
data are time-symmetric then imply {\it the existence of non-trivial
solutions to the vacuum field equations which satisfy the conditions of
definition} (\ref{assidef}) {\it with a finite differentiability of the
rescaled metric which can be chosen to be large. For these solutions 
the rescaled conformal Weyl tensor is differentiable on the conformally
extended space and there exist conformal extensions containing regular
points} $i^{\pm}$ {\it which represent future and past time-like
infinity}.  This resolves a question which remained open and
controversial for forty years.

While it is reassuring to see that we are not discussing a class of
solutions which consists only of Minkowski space, it is clear
that the result does not answer all our questions. The solutions obtained
here are very special. Being Schwarzschildean near space-like infinity,
their NP constants vanish (\cite{friedrich:kannar1},
\cite{newman:penrose:NP1}). As a consequence the rescaled conformal Weyl
tensor $d^i\,_{jkl}$ vanishes at
$i^{\pm}$ (\cite{friedrich:schmidt}, \cite{newman:penrose:NP2}). However,
from the numerical calculation of solutions with a regular $i^+$ we know
that in general $d^i\,_{jkl}$ does not vanish at $i^+$
(\cite{huebner:2001}). This suggests that modifying initial data sets to
become Schwarzschild near infinity considerably reduces their radiation
content. 

Hopefully, there will be developed generalizations of the
method which will allow us to glue other stationary data to given data 
such that we prescribe multipoles moments and non-vanishing NP
constants. Then the argument above will not be available any longer
and the data will gain in interest in the modeling of physical systems. 

But this will still not help us answer the decisive question:
what degree of restriction is implied by `asymptotic simplicity'    
in the class of all (`reasonably' defined) asymptotically flat solutions.

\subsection{The regular finite Cauchy problem}

Under the conformal embedding of Minkowski space into the Einstein cosmos
described above, the Cauchy hypersurface $\tilde{S} = \{t = 0\} \simeq
\mathbb{R}^3$ with its interior Euclidean metric $\tilde{e}$ is mapped
conformally into the unit 3-sphere, $\tilde{S} \rightarrow S \equiv
\{s = 0\} = \tilde{S} \cup \{i\}$, with its standard metric $d\omega^2$
and a point $i$ with coordinates $s = 0$, $\chi = \pi$ which represents
space-like infinity for $(\tilde{S}, \tilde{e})$. If more general data
$\tilde{h}$, $\tilde{\chi}$ satisfying the vacuum constraints are
prescribed on $\tilde{S}$ which are asymptotically Euclidean in the
sense that $\tilde{h}$ suitably approaches the metric $\tilde{e}$ near
space-like infinity, it appears reasonable to conformally compactify
$(\tilde{S}, \tilde{h})$ by suitably rescaling
$\tilde{h} \rightarrow h = \Omega^2\,\tilde{h}$ and adding a point to
$\tilde{S}$ as above. Depending on $\tilde{h}$, there may not exist
coordinates $x^{\alpha}$ on
$S$ near $i$ and a conformal factor $\Omega$ such that the metric
coefficients $h_{\alpha \beta}$ will be smooth at $i$. But even if the data
$\tilde{h}$, $\tilde{\chi}$ are chosen such that $h$ extends smoothly to
$i$, in fact, even if $\tilde{\chi} = 0$, it turns out that in coordinates
adapted to $h$ the rescaled conformal Weyl tensor behaves as
\begin{equation}
\label{funddsing}
d^{\mu}\,_{\nu \lambda \rho} = O(r^{-3}) 
\quad\mbox{as}\quad
r \rightarrow 0,
\end{equation}
where $r$ denotes the $h$-distance from $i$,
unless the ADM mass of the data vanishes, i.e. $\tilde{h}$ is flat. 

To let things not look utterly hopeless, we note that
$d^{\mu}\,_{\nu \lambda \rho}$ is not an arbitrary tensor with Weyl
symmetry satisfying (\ref{funddsing}), that its structure near $i$ is
restricted by the assumption that $\tilde{h}$ and $\tilde{\chi}$ be
asymptotically Euclidean, satisfy the constraints, and that
$d^{\mu}\,_{\nu \lambda \rho}$ is given by a specific formula. We refer to
\cite{friedrich:i-null} for  explicit general expressions. 

In the case of Minkowski space the hypersurfaces ${\cal J}^{\pm}$, which
are swept out by the future resp. past directed null geodesics emanating 
from the point $i^0$ (naturally identified with $i \in S$), are
characteristics for the conformal field equations in the conformally
extended space-time. If we assume for a moment that a similar picture 
can be established for the space-time which develops from non-trivial
data $\tilde{h}$, $\tilde{\chi}$, then (\ref{funddsing}) makes us wonder
why this strong singularity at $i$ does not spread along the
characteristics ${\cal J}^{\pm}$ and destroy the smoothness of the
conformal structure there in the first place ! If such a contradiction does
not arise, it can only be due to (a) specific features of the field
equations and/or (b) specific requirements on the data which are
analogous to the regularity conditions observed in the hyperboloidal
initial value problem. 

If features as referred to in (a) exist, they must lie beyond
conformal regularity. If requirements such as those referred to in (b)
exist  we may wonder what kind of conditions could, in a situation where
the conformal extension is only given by a point, replace the vanishing of
certain functions on a 2-surface as required in the hyperboloidal problem.
Moreover, if one tries to analyze  an initial value problem for the
conformal field equations for data  as singular as
(\ref{funddsing}) the choice of a gauge becomes a delicate problem and
there is little lee-way for rough estimates. 

Somewhat unexpectedly it turns out that for suitably chosen
asymptotically flat initial data {\it the standard Cauchy problem can
be reformulated as a regular finite initial value problem for the
general conformal field equations with data given on a compact
manifold with boundary}.  \index{initial~value~problem!regular~finite}
Moreover, the underlying setting also allows us to analyze to some
extent the questions raised above.

The underdetermined quasi-linear system of constraints imposes very weak
conditions on the data. In particular, it requires very little with
respect to the smoothness of the data at space-like infinity. Thus it
is clear that we have to make assumptions concerning the precise
behaviour of the data near $i$. 
In \cite{friedrich:i-null}, were the setting has been developed and worked
out in detail as described below, it was assumed that the data are
time-symmetric, that $h$ extends smoothly to $i$, and that $h$ is in fact
analytic in normal coordinates at $i$. It follows in particular that
$\tilde{\chi} = 0$ and that the conformal factor $\Omega$ which relates
$h$ to the physical metric $\tilde{h} = \Omega^{-2}\,h$ satisfies
$\Omega \in C^2(\bar{S}) \cup C^{\infty}(\bar{S}
\setminus \{i\})$ with $\Omega = 0$, $d\,\Omega = 0$, $Hess\,\Omega \simeq
h$ at $i$, but $\Omega$ is not smooth at $i$ unless the data have
vanishing ADM mass. As a consequence we still have (\ref{funddsing}).
It should be emphasized that our assumptions have
been made for convenience because the structure of the data near
space-like infinity needs to be known in all details (cf. the remarks
below). In particular, the assumption of analyticity can be dropped and
replaced by a $C^{\infty}$ condition.

\subsubsection{The cylinder at space-like infinity}

In the following we shall describe the main considerations leading to the
formulation of the regular finite initial value problem near space-like
infinity. The general conformal field equations will be used here for
several reasons. We want to obtain a formulation which is leading to
statements which depend as far as possible on the conformal structure
itself and as little as possible on conditions `put in by hand'. 
The nature of the singularity at $i$ forces us to carefully
distinguish between the dependence of the fields on radial and angular
directions at $i$. The frame field in the formalism is readily adapted to
this situation. Much more importantly, we want to avoid any gauge which
depends on implicit wave equations, because these might transport
non-intrinsic singularities along null infinity. An example for this
occurs if the conformal gauge in the metric conformal field
equations is removed by prescribing the Ricci scalar. Clearly, space-like
infinity is not met by time-like conformal geodesics starting at points of 
$\tilde{S} = S \setminus \{i\}$ and conformal Gauss systems should thus be 
unaffected by the singularity at $i$. Finally, the location of the
set where the conformal factor vanishes will be known a priori.

In \cite{friedrich:i-null} the general
conformal field equations have been used in the spin frame formalism,
because this leads to various algebraic simplifications. Here we will
discuss the standard frame formalism, suppressing many details for which
we refer the reader to \cite{friedrich:i-null}. 

Assume a fixed choice of the conformal scaling for the initial data. 
Choose a fixed oriented $h$-orthonormal frame $e_a$, $a = 1, 2, 3$ at $i$.
Any other such frames at $i$ is then obtained in the form 
$e_a(s) = s^c\,_a\,e_c$ with $s = (s^c\,_a) \in SO(3)$. {\it For given $s$,
we distinguish $e_3(s)$ as the radial vector at} $i$ and assume
$e_a(s)$ to be parallelly transported with the Levi-Civita connection of
$h$ along the geodesics with tangent vector $e_3(s)$ at $i$. We denote by
$\rho$ the affine parameter on these geodesics which vanishes at $i$ and
denote the frame obtained from $e_a(s)$ at the value $\rho$ by 
$e_a(\rho, s)$. It will be assumed that $|\rho| < a$, where $a$ is chosen
such that the metric ball $B$ centered at $i$ with radius $a$ is a convex
normal neighbourhood for $h$. Then $]- a, a[ \times  SO(3) \ni (\rho, s)
\rightarrow e_a(\rho, s) \in  SO(S)$ defines a smooth embedding of a
4-dimensional manifold into the bundle $SO(S)$ of oriented
orthonormal frames over $S$.

We denote by $\hat {B}$ the image of the set $[0, a[ \times SO(3)$, by
$I^0$ its boundary $\{\rho = 0\} \simeq SO(3)$, and by $\pi$ the
restriction to $\hat {B}$ of the projection of $SO(S)$ onto $S$. For
given $s \in SO(3)$ the vectors $e_3(s\,s')$ are identical for $s' \in
SO(2)$, the subgroup of $s = (s^c\,_a) \in SO(3)$ for which
$s^c\,_3\,e_c = e_3$, and all these vectors are parallelly transported
along the same geodesic. It follows that the set $\hat {B}$ inherits
from $SO(S)$ an action of the subgroup $SO(2)$ which implies a
factorization $\hat{B} \stackrel{\pi'}{\rightarrow} B' = \hat{B}/SO(2)
\stackrel{\pi''}{\rightarrow} B$ of the map $\pi$ such that $\pi''$
maps the set $\pi'(I^0) \simeq S^2$ onto $i$ while it implies a
diffeomorphism of $B' \setminus \pi'(I^0)$ onto the punctured ball $B
\setminus \{i\}$ which we could use to identify these sets.

It turns out more convenient, however, to pull the initial data on $B$
via $\pi$ back to $\hat{B}$, use this set as the initial manifold and
$\rho$ and $s$ as `coordinates' on it. Questions of smoothness are
then easily discussed and certain expressions simplify
considerably. Since all fields have a well defined transformation
behaviour (spin weight) under the action of $SO(2)$ and this action
commutes with the evolution defined by the propagation equations, no
problem will arise from this. In \cite{friedrich:i-null} it is shown
how to prescribe on the part of $\hat{B}$ where $\rho > 0$ a frame
field $X$, $c_a(\rho, s)$, $a = 1, 2, 3$, where $X$ is generated by
the action of $SO(2)$, the vector fields $c_a(\rho, s)$ satisfy
$T(\pi)c_a(\rho, s) = e_a(\rho, s)$, and a frame formalism can be
applied. In the coordinates $\rho$ and $s$ on $\hat{B}$, which are
essentially spherical coordinates, some frame vector fields and
certain data derived from them are singular at $\rho = 0$.

With a suitable choice of the initial data for the conformal geodesics the
conformal factor and the 1-form $d_k$ take the form 
\begin{equation}
\label{gengau}
\Theta = \frac{\Omega}{\kappa_*}\,\left(
1 - \tau^2\,\frac{\kappa_*^2}{\omega_*^2}\right),\qquad
d_0 = - 2\,\tau\,\frac{\kappa_*\,\Omega_*}{\omega_*^2},\qquad
d_a = \frac{1}{\kappa_*}\,(e_a(\Omega))_*
\end{equation} 
where $\omega^2 = 2\,\Omega\,\,|h^{\alpha \beta}
\Omega_{,\alpha}\,\Omega_{,\beta}|^{-1/2}$ and
$\kappa$ is a function on $\hat{B}$ which we are free to choose. 
Any function $f_*$ in (\ref{gengau}) is considered to be constant along
the conformal geodesics and to agree with its given initial value $f$. 
In the conformal extension of Minkowski
space in which space-like infinity is represented by the point $i^0$ we
would have $\kappa = O(1)$ as $\rho
\rightarrow 0$. For our purpose it will be much more useful to choose
$\kappa$ in the form $\kappa = \rho\,\kappa'$ with some smooth scalar (i.e.
$SO(2)$-invariant) function $\kappa'$ on $\hat{B}$ with $\kappa' = 1$ on
$I^0$ such that $\Theta|_{\tau = 0}  = \frac{\Omega}{\kappa_*} =
O(\rho)$ as $\rho \rightarrow 0$. {\it In this particular
conformal gauge all data extend smoothly to $I^0 \subset \hat{B}$ and the
coefficient functions $\Theta$, $d_k$ in the general conformal field
equations are smooth as well}. 

We are free to choose $\kappa'$. For definiteness we set $\kappa =
\omega$ such that $\Theta = 0$ for $\tau = \pm 1$. If the resulting
coordinates remain regular long enough, future and past null infinity
will be represented near space-like infinity by the sets
${\cal J}^{\pm} = \{\tau = \pm 1\}$, while space-like infinity will now
be represented by the \index{space-like~infinity!cylinder at} {\it
cylinder at space-like infinity} which is given by the set $I = \{\rho
= 0,\,\,|\tau| < 1\}$. It `touches' null infinity at the sets $I^{\pm}
= \{\rho = 0,\,\,\tau = \pm1\}$. What in the beginning was a Cauchy
problem now looks like an initial boundary value problem. It turns
out, however, that the boundary $I$ is of a very special nature.

If we write $w = (e^{\mu}\,_k,\hat{\Gamma}_i\,^j\,_k,\hat{L}_{jk})$ and
the unknown (\ref{gcfeunknown}) in the form $u = (w, z)$ with
$z = (d^{i}\,_{jkl})$, the symmetric hyperbolic reduced equations take
the form 
\begin{equation}
\label{redequ}
A^{\mu}(w)\,\partial_{\mu}\,z = H(w)\,z,\qquad
\partial_{\tau}\,w = F(w,z),
\end{equation}
(where we suppress the dependence of $F$ on the conformal factor $\Theta$
and the 1-form $d_k$).  Extending the data on $\hat{B}$ as well as the
evolution equations (preserving symmetric hyperbolicity) smoothly into the
domain $\rho < 0$, standard results imply the existence of a smooth local
solution in a neighbourhood of the initial hypersurface which contains a
piece of $I$. The fact that $\Theta = 0$ on $I$ then allows us to conclude
from the details of the equations above that
\begin{equation}
\label{arhonull}
A^{\rho} = 0 \quad\mbox{on}\quad I.
\end{equation}
This implies that $I$ is a \index{characteristic!total} {\it total
characteristic} in the sense that the complete system (\ref{redequ})
reduces on $I$ to a symmetric hyperbolic system of inner equations on
$I$. It follows that $u$ is determined uniquely by its values on
$I^0$, that the solution obtained for $\rho \ge 0$ does not depend on
the chosen extension of the data through $I^0$, and that $u$ extends
smoothly to $I$ (which can not be taken for granted in general initial
boundary value problems, cf.  \cite{friedrich:AdS},
\cite{friedrich:nagy}).

%\kappa

The construction of $I$ can be considered as attaching a boundary at
space-like infinity to the physical space-time. However, we wish to
emphasize that not the definition of a boundary but the analysis of the
field equations is our main interest. We do not impose any conditions on
the time evolution of the solutions except the field equations. The
construction is defined in terms of general properties of conformal
geometry, the conformal field equations, and the choice of the initial
data for the gauge. 

While everything extends smoothly to $I$ near $I^0$, it should be observed
that $I$ is not a geometrical entity but that it is attained as a certain
limit of geometric objects, namely the conformal geodesics defining our
gauge. It is not a time-like hypersurface and some of the metric
coefficients do in fact diverge at $I$ in the given coordinates. This is
consistent with the smoothness of the frame coefficients and the
vanishing of the matrix $A^{\rho}$ on $I$. Since the wave equation for the
rescaled conformal Weyl tensor would degenerate at $I$, it turns out
important here that we extract the hyperbolic system from the first order
Bianchi equation. It is remarkable that the blow-up procedure
$i^0 \rightarrow I$, the coordinate $\rho$ which is adapted to the
conformal scaling defined on $S$ by $\Omega$, and the conformal gauge
defined by $\Theta$ conspire to produce a regular finite problem.  By a
simple rescaling (different choice of $\kappa$) and a corresponding
redefinition of the coordinates the representation of space-like infinity
in terms of the cylinder $I$ can be converted into the conventional
representation of space-like infinity by a point $i^0$
(\cite{friedrich:i-null}).

\subsubsection{The transport equations on the cylinder at space-like
infinity}

The procedure described above, by which the point $i^0$ is replaced by the
cylinder $I$, does not only give us regular data and equations but also
leads to an unfolding of the evolution process near space-like infinity
which allows us to analyze the process there at arbitrary order and in all
details. Taking formal derivatives of the evolution equations with respect
to $\rho$, restricting to the cylinder, and observing (\ref{arhonull}),
one obtains transport equations on
$I$ for all the functions
$u^p = \partial^p_{\rho}\,u|_{I}$, $p = 0, 1, \ldots$, on the cylinder. 
Expanding $u^p = u^p(\tau, s)$ in terms of a certain functions system 
$T_{2m}\,^i\,_k$ (where $m = 0, 1, 2, \ldots$ and $i, k = 0, \dots, 2\,m$),
on
$SO(3)$ (resp. on $SU(2)$ in the case of the spin frame formalism used in
\cite{friedrich:i-null}), these equations reduce to a hierarchy of ODEs
along the curves $]-1 , 1[ \ni \tau \rightarrow (\tau, \rho = 0, s) \in
I$, where the differential operator depends on $p$ but not on the data
while the right hand side of the equation for $u^{p+1}$ depends
quadratically on $u^0, \ldots, u^p$ and on the coordinates. 

The calculation of $u$ on $I$ allows us to determine $A^{\mu}(u)$ on $I$,
which gives in particular
\[
A^{\tau} = diag(1+\tau, 1, 1, 1, 1-\tau) \quad\mbox{on}\quad I.
\]
While the second of equations (\ref{redequ}) is an ODE which is
regular at $I^{\pm}$, the matrix $A^{\tau}$ occurring in the first of
equations (\ref{redequ}) degenerates precisely on $I^{\pm}$.
{\it All the remaining open questions concerning the regular finite
initial value problem near space-like infinity are related to this fact}.

This degeneracy (which is removed at the next order in $\rho$) has
important consequences. {\it It turns out that in general the quantities
$u^p$, in particular the expansion coefficients of the rescaled conformal
Weyl tensor, contain terms which behave as
$(1 - \tau)^k\,\log^j(1 - \tau)$ at $I^+$}. In the expressions which so
far have been calculated explicitly only terms with the powers $j =
0, 1$ occur, but due to the non-linearity of the equations other powers
should arise as well.  Because of the hyperbolicity of the reduced
equations these singularities can be expected to spread along ${\cal
J}^{\pm}$. The powers $k$ and $j$ observed in the explicit calculations
then suggest that in general not all components of $d^i\,_{jlm}$ will be
bounded at null infinity. In fact, in the spinor notation used in
(\ref{bianrewr}), (\ref{Pcond}) it appears that in general the rescaled
Weyl spinor field will have an unbounded component
$\phi_{abcd}\,\iota^a\,\iota^b\,\iota^c\,\iota^d$ which behaves as 
$\log (1 - \tau)$ at ${\cal J}^+$, where $\tau \rightarrow 1$. 

This suggests that in general the conformal Weyl tensor will vanish on
${\cal J}^+$ but will neither be differentiable there nor satisfy the
assumption (\ref{Pcond}).

Some care is required here, because the hyperbolicity does not extend to
$I^{\pm}$. However, if the whole setting developed in
\cite{friedrich:i-null} is linearized at  Minkowski space, a complete
analysis of the situation can be given which confirms the conclusion (cf.
J. Valiente-Kroon, this volume).  

Due to the geometric nature and the conformal invariance of our gauge
conditions we are discussing here intrinsic features of the solutions. If
the null infinities of the solutions would admit smooth extensions of the
conformal structures, then the conformal geodesics, the conformal factor,
and the frame would extend smoothly through null infinity.  Also the
components of the rescaled conformal Weyl tensor in that frame would
then extend as smooth functions of $\tau$ through null infinity. 
Thus the observations above suggest that {\it even under assumptions on
the  conformal data which cannot be improved from the point of view of}
smoothness {\it there will in general occur logarithmic singularities
at null infinity due to the evolution process}.

\subsubsection{A conjecture}

The class of logarithmic singularities determined in
\cite{friedrich:i-null} have been found by solving the transport
equations on $I$ explicitly for certain components ${u'}^{p}$ of the
functions $u^p$, $p = 0, 1, 2, \ldots$, in the decomposition with
respect to the functions $T_{2m}\,^i\,_k$. Inspecting the initial data
${u'}^{p}_*$ for these quantities on $I^0$ and relating them to the
free datum $h$ on $\tilde{S}$, we find that the logarithmic terms we
find for general data in our class do not occur in ${u'}^{p}$ for $p
\le q_* + 2$ if and only if $h$ satisfies the
\index{space-like~infinity!regularity~condition~at} {\it regularity
condition}
\begin{equation}
\label{hf:regcond}
{\cal S}(D_{i_1}, \dots, D_{i_q}\,B_{jk}(i)) = 0
\quad\mbox{for}\quad q = 0, 1, \ldots, q_*,
\end{equation}
where $D$ denotes the connection and  
$B_{jk} = \frac{1}{2}\,\epsilon_j\,^{il}\,D_i\,L_{lk}$ the Cotton tensor
of $h$ while ${\cal S}$ denotes the operation of taking the symmetric
trace-free part of a tensor. 

Consistent with the fact that the free initial datum is given by the
conformal structure of $h$, condition (\ref{hf:regcond}) is conformally
invariant for given $q_* \ge 0$. The condition has been found earlier in
the quite different, though related, context of vanishing ADM mass
(\cite{friedrich:static}). Although the derivation is much easier in that
case and related to a geometric picture, we are still lacking a
geometrical interpretation of condition (\ref{hf:regcond}).

Since the condition is conformally invariant and it is easy to construct
data which are conformally flat near space-like infinity, there exist many
non-trivial data satisfying the condition. With the techniques of
\cite{chrusciel:delay} and \cite{corvino} such data can even be
constructed so that they agree with prescribed data on a compact set.
Moreover, there exists a large class of data which satisfy the condition
in a non-trivial way (\cite{friedrich:i-null}). In fact, all data which
are conformal to static data near space-like infinity satisfy
(\ref{hf:regcond}) for all $q_*$ (\cite{friedrich:static}). Since static
data are analytic near space-like infinity (\cite{beig:simon},
\cite{kennefick:o'murchadha}), it follows that condition (\ref{hf:regcond})
can be satisfied with Cotton tensors which do not vanish near space-like
infinity (\cite{friedrich:i-null}). This is consistent with the fact that
the static solutions represent the largest class of time-symmetric
solutions which are known to admit near space-like infinity a smooth
structure at null infinity. It would be interesting to know whether the
results of \cite{chrusciel:delay} and \cite{corvino} can be generalized
to glue static ends to given data.

For various reasons we expect condition (\ref{hf:regcond}) to be of a much
wider significance.

{\bf Conjecture}: {\it There exists an integer $k_* \ge 0$ such that for
given 
$k \ge k_*$ the time evolution of an asymptotically Euclidean,
time-symmetric, conformally smooth initial data set admits a conformal
extension to null infinity of class $C^k$ near space-like infinity, if the
Cotton tensor satisfies condition} (\ref{hf:regcond}) {\it for a certain
integer $q_* = q_*(k)$}.

If this conjecture can be shown to be correct, solutions for data which
satisfy  (\ref{hf:regcond}) for all $q_*$ will contain smooth
hyperboloidal hypersurfaces. We expect that the corresponding
hyperboloidal initial data approach Minkowskian hyperboloidal data in a
continuous way if the Cauchy data approach Minkowski data. This then can
be used in conjunction with the results of \cite{friedrich:n-geod}
about the hyperboloidal Cauchy problem to show the existence of solutions
satisfying all the requirements of definition (\ref{assidef}). Because of
the degeneracy of the equations at $I^{\pm}$ the proof of the conjecture
is not a standard problem and requires new ideas. However, we can split
off an important and interesting subproblem.

{\bf Subconjecture}: {\it If condition} (\ref{hf:regcond}) {\it holds for
a given  $q_* \ge 0$, the functions $u^p$, $p \le q_* + 2$, extend
smoothly to $I^{\pm}$}. 

In the formalism employed in \cite{friedrich:i-null} the validity of the
statement above can in principle be checked by a recursive calculation of
the functions $u^p$. In fact, such a calculation shows that the $u^p$
are smooth at $I^{\pm}$ for $p \le 3$ if the corresponding regularity
conditions hold (\cite{friedrich:i-null}, \cite{friedrich:kannar1}). 
The general proof essentially reduces to controlling the algebraic
structure of the terms on the right hand side of the ODEs governing the
transport of the $u^p$, which, with increasing order, becomes quite
complicated (cf. the discussion in
\cite{friedrich:kannar1}, \cite{friedrich:kannar2}).
It is most desirable to determine terms of higher order by an explicit
calculation, possibly with an algebraic computer program, to obtain
an insight into the general structure of the functions $u^p$. This will
help with a general analytic proof of the subconjecture and also provide
indispensable information for solving the regular finite initial
value problem numerically.   

\subsubsection{Problems and prospects}

Condition (\ref{hf:regcond}) suggests that data for solutions satisfying
the requirements of definition (\ref{assidef}) have to satisfy at $i$
conditions at all orders. 
Therefore, a generalization of the regularity condition to the
case of data with non-trivial extrinsic curvature requires a general and
very detailed investigation of the solutions to the constraints near
space-like infinity. Such a study has recently been carried out (cf. S.
Dain, this volume)  and the generalization of the analysis of
\cite{friedrich:i-null} to a much larger class of data, which also 
requires new considerations concerning the gauge, is possible.

The proof of our conjecture, or its generalization to the case of
non-trivial extrinsic curvature, requires to show the existence of
solutions to the finite regular initial value problem near
$I$ whose underlying manifolds contain a piece of the form 
$\{|\tau| < 1, 0 \le \rho < \rho_0\}$. 
Moreover, we need to obtain sufficient control on the smoothness of
the solutions near 
${\cal J}^{\pm}_{\rho_0} = \{|\tau| = 1, 0 \le \rho < \rho_0\}$. The
specification of the precise values of $k_*$ and $q_*(k)$, possibly
referring to Sobolev spaces, should be part of this proof. In principle it
is not only of interest to see whether the conjecture can be justified but
we would also like to know how  the logarithmic terms propagate along
${\cal J}^+_{\rho_0}$, if they do so, and what the role and meaning of
these terms is.

It would also be desirable to sort out for more general classes
of free data the type of non-regularities which arise from solving the
constraints and from solving the evolution equations. Since free data can
be quite `rough' at space-like infinity while still being asymptotically
Euclidean, there has to be made a reasonable choice. There is no point in
striving for a generality which is not required and of no use in modeling 
physical systems. However, already for quite natural looking free data the
constraints can imply  terms of the form $\rho^k\,\log^j\rho$ on
$\tilde{S}$ (cf. S. Dain, this volume). Clearly it is of interest to see
to what extent our analysis can be generalized to handle such data even if
our conjecture and its possible generalizations to more general data bear
out. It will be the only way to assess the generality and the role of the
solutions which admit a smooth structure at null infinity in the class of
all solutions arising from asymptotically Euclidean data. 

If the smoothness of the conformal structure obtained at null infinity
turns out to be much weaker than expected in systems which appear of
`physical importance'\footnote{Besides getting control on the
mathematical possibilities, finding out which data are of
`physical importance' may turn out to be the most interesting part of the
whole exercise. The interpretation of initial data is still a wide open
field.}, we will have to reconsider definition (\ref{assidef}). 
While it may not be applicable any longer in
the desired generality, the  underlying idea is viable because the
conformal field equations may be used to control also solutions with a
`rough' asymptotic behaviour  (\cite{chrusciel:lengard}, \cite{lengard},
cf. also P. T. Chru\'sciel, this volume). If the solutions for suitable
data admit polyhomogeneous expansions in
$\tau$, we can expect to relate the coefficients of this expansion, which
can be considered as functions on ${\cal J}^+$, again to the initial data.
This should allow us to give some interpretation for these terms.   

In all such studies the insight into the equations gained by trying to
establish the extremely sharp characterization of the fall-off
behaviour of gravitational fields in terms of asymptotic
simplicity\index{asymptotic!simplicity} will prove important.  For
instance, the general analytic proof of the subconjecture stated above
will be of particular interest because it will provide a further
understanding of the peculiar features of the field equations.  The
precise structure of the lower order terms in the equations, which is
essentially determined by the fact that we are dealing with a soldered
gauge theory (cf. the discussion of the general conformal field
equations in the context of normal conformal Cartan connections in
\cite{friedrich:AdS}), becomes critical here.  Understanding its role
in the context of the data dependent conformal singularity at
space-like infinity may shed light also on the mechanism of the field
equations in other singular situations.

The concept of an isolated system is not part of the general theory but
a matter of expediency (cf. the discussion in
\cite{friedrich:roger}). The final criterion for its definition is the
possibility to model physical systems in a way which allows us to extract
interesting information about physical processes. If the conjecture above
bears out, the resulting space-times will allow us to perform very
detailed discussions of all their physical aspects and the proof of the
conjecture will provide information which reduces many discussions to
direct calculations. 

An example for this is given by the calculation of the NP-constants in
terms of the initial data in \cite{friedrich:kannar1}. As neatly
discussed in \cite{dain:valiente-kroon}, these provide interesting 
restrictions on the global structure of solutions. Another example is the
decision on the reducibility of the BMS group to the Poincar\'e group in
\cite{friedrich:kannar1}. So far both results still rely on the
correctness of our conjecture. In fact, many earlier studies of physical
concepts associated with the asymptotic structure will be put on a firm
analytical footing. 

If logarithmic singularities at null infinity (or the coefficients of the
logarithmic terms) turn out to be of any physical significance this will
not only create problems for the concept of asymptotic simplicity but
also for the conventional method of calculating gravitational radiation
by solving numerically initial boundary value problems for Einstein's
field equations. In that approach the singularities and what might be
lost by ignoring them cannot even be seen. While the initial data for the
initial boundary value problem can be arranged not to be affected by the
modification of the data near space-like infinity, with our present
knowledge there appears to be no way, whatever boundary treatment in the
numerical calculation is used, to control whether one is calculating a
solution where the influence of such a modification enters through the
boundary into the numerical solution (cf. \cite{friedrich:nagy} for the
amount of freedom and control available in the analytic treatment of the
initial boundary value problem for Einstein's field equations).

On the other hand if the latter method makes any sense at all, it may for
practical purposes be quite sufficient and of advantage to calculate
numerically asymptotically simple solutions for data with a Schwarzschild
end {\it if the procedures in} \cite{chrusciel:delay}, \cite{corvino}
{\it are amenable to a numerical treatment}. It is quite simple to
identify in the evolution of such data hyperboloidal hypersurfaces which
intersect ${\cal J}^+$ in the Schwarzschild part of the evolution and
which extend with the implied data smoothly as solutions to the conformal
constraints across ${\cal J}^+$ (\cite{friedrich:cg on vac}). In fact, the
corresponding hyperboloidal data for the conformal field equations can
be calculated by solving a system of ODEs (\cite{friedrich:i-null}).  
This would resolve a number of difficult problems occurring in the present
numerical codes (cf. J. Frauendiener and S. Husa, this volume). 

Again, while this in the end may turn out sufficient for practical
purposes, we have no possibility to decide on that before the existence of
asymptotically simple solutions has been discussed under sufficiently
general assumptions. In that case quite general complete maximal
hyperbolic solutions together with their asymptotics may become
calculable from standard Cauchy data and the interpretational framework
associated with the concept of asymptotic simplicity may allow us to draw
the desired conclusion. 

It appears that presently the conformal field equations offer the only way
to perform such calculations. In the case of the characteristic method,
which also allows one to calculate numerically semi-global solutions,
there may arise problems because null hypersurfaces have the tendency to
develop caustics (an intrinsic feature which cannot be overcome by
smoothing procedures, cf. \cite{friedrich:stewart} for a discussion of
caustics in the context of the characteristic initial value problem and
G. Galloway, this volume, for an illustration of the ubiquity of
conjugate points on null geodesics). There are also problems in
prescribing appropriate initial data. Assuming past time-like infinity to
be represented by a regular point, one could think of prescribing null
data on the set ${\cal J}^- \cup \{i^-\}$ which only has a very
special and well understood `caustic' at $i^-$. However, at present it is
by no means clear how to prescribe these data such that they develop into
a solution which admits an asymptotically flat Cauchy hypersurface.
Similar difficulties occur if one tries to invent other ways of
calculating complete space-times from characteristic data.

\subsection{Time-like infinity}

It is a remarkable feature of the conformal Einstein equations, which
should not be taken for granted, that they force the null generators of
${\cal J}^+$ to meet in precisely one regular point $i^+$ of the conformal
extension if the solution exists for long enough in a sufficiently
regular fashion and the initial data are prescribed appropriately
(\cite{friedrich:n-geod}). In such a situation `time-like infinity' has a
clear-cut meaning, it is locally similar to that of Minkowski space, and
it can `readily' be attained in numerical calculations
(\cite{huebner:2001}, cf. also S. Husa, this volume). The calculation of
$i^+$ provides an opportunity to test numerical codes in long range
calculations of space-times and to perform studies of the non-linear
interaction of gravitational waves under well controlled conditions,
however, it is of course not a purpose in itself.

To calculate the gravitational radiation emitted by the merger of two
black holes we will have to consider solution which develop singularities
and horizons. While it may appear wise in a numerical calculation to avoid,
if possible, entering the horizon and hitting the singularity, such an
attitude does not help in an analytical investigation and we do, in fact,
consider it necessary to attack the situation head on. In fluid mechanics
there exists a long tradition of studying singularities (shocks) and there
have been developed methods to handle them in numerical calculations.
The singularities observed in general relativity, however, are in general
much more complicated and severe from a conceptional point of view as well
as from the point of view of actual analytical expressions. In spite of a
considerable effort to come to grips with them analytically and, more
recently, numerically (cf. D. Garfinkle, this volume, and also
\cite{berger} and the literature given there, which discusses some
remarkable achievements), we are still far from a situation where we could
attain the desired information in sufficiently general situations. In
fact, the `well-known' Schwarzschild-Kruskal or Kerr solutions still
pose  difficulties. 

There exist several ways to discuss `singular points of
space-times' in an abstract way (cf. \cite{penrose:chandra} and the
references given there), however, there are no concepts available which
would allow us to perform under general assumptions an analysis of
singularities in the context of the field equations. 
While it is a central problem in singular situations to obtain
detailed statements which are independent of any specific gauge, it is at
least as difficult to develop gauge conditions which allow us to arrive
at any interesting statement at all. The latter problem certainly does not
have an universal solution. The best we can hope for is to find manageable
conditions which by their general nature and by experience with specific
solutions may be expected to work for certain classes of solutions. 

We have seen above that conformal Gauss systems have many nice local
properties and the discussion of the regular finite Cauchy problem suggests
that they also behave nicely near space-like infinity. It can also be shown
that they can be used in calculations of $i^+$ from hyperboloidal Cauchy
data which are sufficiently close to Minkowskian hyperboloidal data.
However, in all these situations the gravitational field must be
considered as weak and one would like to know whether conformal Gauss
systems can also exist globally and with good asymptotic properties on
asymptotically flat space-times which posses regions of strong curvature
or even singularities and horizons. 

Caustics of congruences of metric geodesics arise because the latter
obey equations of second order. Congruences of conformal geodesics are
governed by equations of third order (cf. \cite{ogiue}) and are thus
even more prone to caustic formation, in fact, different conformal
geodesics may become tangent to each other. However, the same structural
property which may be responsible for such severe caustics allows us to
show   that {\it there exist conformal Gauss systems on the
Schwarzschild-Kruskal solution which smoothly cover the entire space-time
and extend smoothly and without degeneracy through ${\cal J}^{\pm}$ into
any smooth extension of the space-time through null infinity}
(\cite{friedrich:cg on vac}). 

These systems are characterized by initial data for the conformal
geodesics on the hypersurface $\tilde{S}$ of the Schwarzschild-Kruskal
solution which extends analytically the hypersurface $\{ t = 0 \}$ of the
Schwarzschild space. The data considered in \cite{friedrich:cg on vac}
are invariant under the space reflection symmetry in
$\tilde{S}$ and the  rotational symmetry and the congruence is orthogonal to
$\tilde{S}$. On $\tilde{S}$ are given the standard spherical coordinates
$\phi$, $\theta$ and the isotropic radial coordinate $r$, $0 < r < \infty$,
which is related to the radial coordinate $\bar{r}$ of the standard
Schwarzschild line element  $g_S = (1 - \frac{2\,m}{\bar{r}})\,dt^2
- (1 - \frac{2\,m}{\bar{r}})^{- 1}\,d\bar{r}^2 - \bar{r}^2\,d\,\sigma^2$
on the
subset $\bar{r} < 2\,m$ by 
$r = 1/2\/(\bar{r} - m + \sqrt{\bar{r}\,(\bar{r} - 2\,m)})$. The two
asymptotically flat ends of $\tilde{S}$ are at $r = 0$ and $r = \infty$. The
coordinates $\phi$, $\theta$, $r$ are dragged along with the congruence
and define together with the parameter $\tau$ of the conformal geodesics a
smooth coordinate system (ignoring here the singularity inherent in the
spherical coordinates which in the present situation can be easily 
avoided) on the Schwarzschild-Kruskal space with $\tilde{S} = \{\tau =
0\}$. The initial data are prescribed such that the conformal factor
takes the form 
$\Theta = (r - \frac{m}{2})^2(r + \frac{m}{2})^{-2}(f(r)^2 - \tau^2)$ 
where $f(r) \equiv r/(r^2 - \frac{m^2}{4})$.

The coordinates stay regular, because of the specific choice of the 1-form
$b$ on $\tilde{S}$. The acceleration of the conformal geodesics,
which is determined by $b$, tends by our choice to spread out the curves
such that no caustics can develop. This latter fact is established by
analyzing an equation satisfied by the connecting vectors of the congruence
of conformal geodesics which generalizes the Jacobi equation known for
metric geodesics. We consider the fact, that such a geometric analysis is
possible as particularly useful.

It is well known that the Schwarzschild-Kruskal solution admits global
double null coordinates (cf. B. Schmidt, this volume), however, these
coordinates, or coordinates derived from them by simple algebraic relations,
do not extend smoothly through ${\cal J}^{\pm}$ and they are very specific
for the Kerr family. The properties of the conformal Gauss systems mentioned
above raise the hope that such systems can be constructed in much more
general situations. Of course, while there are a few general considerations
which may help here, a good choice of the initial data for the conformal
geodesics will depend on the specific solution and one has to gather
experience with this, as with any choice of gauge. If conformal Gauss systems
are used in the context of the general conformal field equations to perform
numerical calculations, the tensor fields entering the generalized Jacobi
equations are essentially calculated as one goes along and the possible
tendency of the underlying congruence of conformal geodesics to form
caustics can be monitored during the calculation.

It should be noted that Gauss systems for which the initial tangent vector
field on $\tilde{S}$ and the pull back of the 1-form field $b$ to
$\tilde{S}$ are identical, define the same congruence of curves, if these
are considered as point sets. The parameterizations of these curves, however,
may be different and the systems will thus  define different foliations.
This freedom has hardly been explored yet.

Since the space reflection in $\tilde{S}$ is given by $r \rightarrow
m^2/4\,r$, it is sufficient to discuss the conformal geodesics starting at
the subset of $\tilde{S}$ where $r \ge m/2$, which cover a full $\bar{r} \ge
2\,m$ part and two `halves' (one in the past and one in the future of
$\tilde{S}$) of the $\bar{r} < 2\,m$ part of the Schwarzschild solution. The
conformal geodesics with 
$\phi = \mathrm{const}.$, $\theta = \mathrm{const}.$, $r = r_+ \equiv (3 + \sqrt{5})\,m/4$,
stay on a hypersurface $\{\bar{r} = 5\,m/2\}$ of the
Schwarzschild-Kruskal space and approach time-like infinity, denoted by
$i^+$, as $\tau \rightarrow
\tau_i = 2/\sqrt{5}\,m$. The conformal geodesics starting with $r > r_+$ 
run out to null infinity and arrive there as 
$\tau \rightarrow \tau_{scri} = f(r)$, which is consistent with the
expression above of the conformal factor.

The curves starting with $\frac{m}{2} \le r < r_+$ hit the singularity
at $\bar{r} = 0$ as $\tau \rightarrow \tau_{\mathrm{sing}} \equiv
f(r)\,H(K(k), k)$, where $k(r) \equiv \sqrt{2}\,\left(1 +
\sqrt{m\,r/2}\,(r - \frac{m}{2})\right)^{- 1/2}$, $K(k) \equiv
\int_0^{\pi/2} (1 - k^2\,\sin^2 x)^{- 1/2}\,d\,x$, and
\[
H(x, k) \equiv \tanh \left(
\sqrt{\frac{2 - k^2}{2}}\int_0^x 
\,\frac{dn^2 (v, k) - (1 - k^2)}{dn^2 (v, k) + (1 - k^2)}\,\,d\,v
\right), 
\]
with Jacobi's elliptic function $dn$. They pierce the horizon
${\cal H}^+ = \{\bar{r} = 2\,m\}$ at 
$\tau = \tau_{\mathrm{hor}} \equiv f(r)\,H(x_h, k)$, with
$x_h \equiv \int_0^{\arcsin(k^2/(4-3k^2))} (1 - k^2\,\sin^2 x)^{-1/2}\,d\,x$. 
As to be expected, we find that $\tau_{scri} \rightarrow \tau_i$ as 
$r \rightarrow r_+$ with $r > r_+$ and also 
$\tau_{\mathrm{sing}}, \tau_{\mathrm{hor}} \rightarrow \tau_i$ 
as $r \rightarrow r_+$ with $r < r_+$.

Signals periodic in proper time of an observer approaching the
horizon are received on ${\cal J}^+$ at Bondi-times which grow only
logarithmically\footnote{I owe much to discussions with P. H\"ubner
about this point. However, we both have to acknowledge the priority of W.
Allen's formula: {\it eternity is very long, especially near the
end}.}. Consequently, a high resolution is required in a numerical
calculation which aspires to probe the field near $i^+$ by following
outgoing null geodesics which start on a Cauchy hypersurface and arrive at
null infinity at very late Bondi-times. Therefore we consider it
particularly important to acquire a detailed understanding of the
characteristic features of the field near $i^+$. 

It is customary, and natural from the point of view of the causal
structure (cf. \cite{penrose:chandra}), to denote time-like infinity in
the schematic (conformally compactified) causal diagrams by one or several
ideal points $i^+$, but the specific structure of the field in the
region close to them is little understood. In our representation of the
Schwarzschild-Kruskal space the behaviour of the term
$(\Theta\,\bar{r})^2\,d\,\sigma^2$ of the metric $\Theta^2 g_S$ in the
limit as $\tau \rightarrow \tau_i$ on the hypersurface
$\{\bar{r} = \hat{r}\}$ or $\{\bar{r} = 2\,m\}$ suggests that we
should consider $i^+$ as a point with coordinates $\tau = \tau_i$, $r =
r_+$ (the coordinates $\phi$ and $\theta$ loosing their meaning there as
at the origin of a polar coordinate system). Without further analysis,
however, this can only be considered as a way of speaking. Contrary to
what is observed in Minkowski space, $i^+$ is not only approached by
time-like geodesics and by the null generators of ${\cal J}^+$, but also
by null geodesics originating from the interior of the space-time, namely
those generating the event horizon and those which stay on the set 
$\{\bar{r} = 3\,m\}$ or approach it in the future. 
Such phenomena do not occur for regular conformal structures and thus
$i^+$ cannot be considered as a point in a suitably defined regular
conformal extension.

It is reasonable to expect that conformal geodesics of a regular
conformal structure which approach a point $p$ in the underlying
manifold $M$ can be extended, perhaps after an admissible
reparametrisation, through that point as a smooth solution to the
conformal geodesic equation, and that the curve will be time-like at
$p$ if it was time-like before (while such a statement is known to
hold for metric geodesics it still has to be shown for conformal
geodesics). Assuming this to be true, we can observe an effect of the
conformal singularity near $i^+$ in our global conformal Gauss
system. Since $\partial\,\tau_{scri}/\partial\,r$ converges to a
finite negative number as $r \rightarrow r_+$ with $r > r_+$, we see
that ${\cal J}^+$ approaches, as to be expected, the hypersurface $\{r
= r_+\}$ near $i^+$ under a certain angle. In contrast, we find that
$\partial\,\tau_{\mathrm{sing}}/\partial\,r \approx (r_+ - r)^{-
\frac{\sqrt{2} - 1}{\sqrt{2}}}$ as $r \rightarrow r_+$ with $r < r_+$,
such that the graph of $\tau_{\mathrm{sing}}$, and with it also the
graph of $\tau_{\mathrm{hor}}$, becomes tangent to $\{r = r_+\}$ at $i^+$. As a
consequence, the conformal geodesics generating $\{r = r_+\}$ become
tangent to the null geodesics generating ${\cal H}^+$ when they
approach $i^+$.

Though this may look like a subtlety, one may wonder about its consequences
if one tries to observe the CFL condition in a numerical calculation near
$i^+$. Moreover, the speed with which the singularity and the horizon
approach the hypersurface $\{r = r_+\}$ may require special measures to
sufficiently resolve these structures numerically. 

There are of course much more drastic effects. While the rescaled
conformal Weyl tensor (expressed in suitably adapted coordinates or frame
fields) extends smoothly to ${\cal J}^+$, it diverges along the null
generators of the horizon like $\Theta^{-1}$ near $i^+$, and blows up like
$1/\bar{r}^{3}$ near the singularity (if the conformal factor approaches
a positive value there, as it does in our case), resulting in a 
non-uniformity with strong gradients near $i^+$. Moreover, slight changes
in the data may lead to a Kerr solution with drastically different
behaviour near $i^+$, including a Cauchy horizon, a completely different
singularity structure, and causality violations (cf.
\cite{carter}). In recent years considerable work has been done to
understand what happens to these structures under perturbations of the
data (cf. \cite{israel} and the references given there) but these studies
supply surprisingly little precise information on the behaviour of the
field in the close vicinity of $i^+$.

One may object to our discussion that we create unnecessary problems
by insisting on conformal rescalings in a region where the field is
usually considered as weak. But the phenomena we discuss are present
anyway, they are just brought out most succinctly by suitable
rescalings. It is not so clear, though, what `suitable' may mean
here. In the case of space-like infinity we have seen that there are
representations which allow us to see properties which are hard to
analyse with other choices of the gauge.

This raises the question whether there exists a kind of `resolution of the
singularity at time-like infinity' which makes the situation more amenable to
a detailed analysis and to a numerical treatment. The situation near $i^+$ is
clearly quite different from that near $i^0$. There is nothing near
$i^0$ which compares with the singularity `$\{\bar{r} = 0\}$' and the horizon
${\cal H}^+$ and the way they meet with ${\cal J}^+$ at $i^+$. Because
of these structures the singularity obtained in a conformal representation
which makes time-like infinity a point is much less isotropic than the
singularity obtained if space-like infinity is represented by a point.
The singularity at $\bar{r} = 0$ cannot be removed by a simple
rescaling without shifting it to infinity. Moreover, $i^+$ is causally
differently related to the physical space-time than $i^0$. A `resolution of
the singularity at $i^+$' must be compatible with a time evolution which
essentially takes into account the complete information given by the initial
data. It should be emphasized that we are not interested here in a formal
definition of a space-time boundary which relies on more or less implicit
assumptions of the time evolution of the field. We seek a construction
which will help us analyse the latter. This construction can only be
fruitful if it follows from the structure of the field equations in a
similar way as the cylinder at space-like infinity does.  

Looking again at the complicated behaviour of the fields near time-like
infinity suggested in \cite{israel}, this task may appear hopeless.
However, it seems to be the accepted view that in the case of a gravitational
collapse to a black hole some kind of universal behaviour can be expected
near $i^+$ in the sense that the field will in general approach near
time-like infinity somehow that of a Kerr solution (\cite{penrose:nuovo}).
The recent proof of the Penrose inequality (\cite{bray}, 
\cite{huisken:ilmanen}) lends some support to this, because the original
argument leading to this inequality relied on the `establishment viewpoint'
indicated above (\cite{penrose:inequ}).

%%% Local Variables: 
%%% mode: latex
%%% TeX-master: "lnp"
%%% TeX-master: "lnp"
%%% TeX-master: "lnp"
%%% End: 

\end{document}